# A hierarchy of energy- and flux-budget (EFB) turbulence closure models for stably stratified geophysical flows

S. S. Zilitinkevich [1,2,3,4,5], T. Elperin [6], N. Kleeorin [6], I. Rogachevskii [6] and I. Esau [3]

[1]Finnish Meteorological Institute, Helsinki, Finland
[2]Division of Atmospheric Sciences, University of Helsinki, Finland
[3]Nansen Environmental and Remote Sensing Centre / Bjerknes Centre for Climate Research, Bergen, Norway
[4]Department of Radio Physics, N.I. Lobachevski State University of Nizhniy Novgorod, Russia
[5]A.M. Obukhov Institute of Atmospheric Physics, Moscow, Russia
[6]Department of Mechanical Engineering, Ben-Gurion University of the Negev, Beer-Sheva, Israel

## Abstract

In this paper we advance physical background of the energy- and flux-budget turbulence closure based on the budget equations for the turbulent kinetic and potential energies and turbulent fluxes of momentum and buoyancy, and a new relaxation equation for the turbulent dissipation time-scale. The closure is designed for stratified geophysical flows from neutral to very stable and accounts for the Earth rotation. In accordance to modern experimental evidence, the closure implies maintaining of turbulence by the velocity shear at any gradient Richardson number $Ri$, and distinguishes between the two principally different regimes: "strong turbulence" at $Ri \ll 1$ typical of boundary-layer flows and characterised by the practically constant turbulent Prandtl number $Pr_T$; and "weak turbulence" at $Ri > 1$ typical of the free atmosphere or deep ocean, where $Pr_T$ asymptotically linearly increases with increasing $Ri$ (which implies very strong suppression of the heat transfer compared to the momentum transfer). For use in different applications, the closure is formulated at different levels of complexity, from the local algebraic model relevant to the steady-state regime of turbulence to a hierarchy of non-local closures including simpler down-gradient models, presented in terms of the eddy-viscosity and eddy-conductivity, and general non-gradient model based on prognostic equations for all basic parameters of turbulence including turbulent fluxes.

# Symbols

| | |
|---|---|
| $A_i = E_i / E_K$ | share of the $i$-component, $E_i$, of turbulent kinetic energy, $E_K$ |
| $E = E_K + E_P$ | total turbulent energy (TTE) |
| $E_K = \tfrac{1}{2} \langle u_i u_i \rangle$ | turbulent kinetic energy (TKE) |
| $E_i$ | longitudinal ($i=1$ or $i=x$), transverse ($i=2$ or $i=y$) and vertical ($i=3$ or $i=z$) components of TKE |
| $E_\theta = \tfrac{1}{2} \langle \theta^2 \rangle$ | "energy" of potential temperature fluctuations |
| $E_P$ | turbulent potential energy (TPE), Eq. (28) |
| $F_i = \langle u_i \theta \rangle$ | turbulent flux of potential temperature |
| $F_z$ | vertical component of $F_i$ |
| $f = 2\Omega \sin\varphi$ | Coriolis parameter |
| $\mathbf{g}$ | acceleration due to gravity |
| $K_M$ | eddy viscosity, Eq. (43) |
| $K_H$ | eddy conductivity, Eq. (44) |
| $K_D$ | eddy diffusivity |
| $L$ | Obukhov length scale, Eq. (41) |
| $l$ | turbulent length scale |
| $N$ | mean-flow Brunt-Väisälä frequency |
| $P$ | mean pressure |
| $P_0$ | reference value of $P$ |
| $p$ | fluctuation of pressure |
| $Pr = \nu / \kappa$ | Prandtl number |
| $Pr_T$ | turbulent Prandtl number, Eq. (45) |
| $Q_{ij}$ | correlations between fluctuations of pressure and velocity-shear, Eq. (15) |
| $Ri$ | gradient Richardson number, Eq. (3) |
| $Ri_f$ | flux Richardson number, Eq. (40) |
| $R_\infty$ | maximal $Ri_f$ in homogeneous sheared flow |
| $S = \lvert \partial \mathbf{U} / \partial z \rvert$ | vertical shear of the horizontal mean wind |
| $T$ | absolute temperature |
| $T_0$ | reference value of absolute temperature |
| $t_T = l\, E_K^{-1/2}$ | dissipation time scale |
| $\mathbf{U} = (U_1, U_2, U_3)$ | mean wind velocity |
| $\mathbf{u} = (u_1, u_2, u_3)$ | velocity fluctuation |
| $\beta = g / T_0$ | buoyancy parameter |
| $\gamma = c_p / c_v$ | ratio of specific heats at constant pressure and constant volume |
| $\varepsilon_K$, $\varepsilon_\theta$, $\varepsilon_i^{(F)}$ and $\varepsilon_{ij}^{(\tau)}$ | dissipation rates for $E_K$, $E_\theta$, $F_i^{(F)}$ and $\tau_{ij}$ |
| $\varepsilon_{i3(\text{eff})}$ ($i=1,2$) | effective dissipation rates for the vertical turbulent fluxes of momentum |
| $\kappa$ | temperature conductivity |
| $\nu$ | kinematic viscosity |
| $\Pi = E_P / E_K$ | energy stratification parameter [Eq. (74)] |
| $\Phi_K$, $\Phi_\theta$ and $\Phi_F$ | third-order turbulent fluxes of TKE $E_K$, and the fluxes $E_\theta$ and $F_i$ |
| $\varphi$ | latitude |
| $\tau_{ij}$ | Reynolds stresses (components of turbulent flux of momentum) |
| $\tau_{\alpha 3}$ ($\alpha = 1,2$) | components of the Reynolds stresses representing vertical turbulent flux of momentum |
| $\tau$ | modulus of the horizontal vector ($\tau_{13}, \tau_{23}$) |
| $\rho$ | mean density |
| $\rho_0$ | reference value of $\rho$ |
| $\Theta$ | mean potential temperature |
| $\theta$ | fluctuation of potential temperature |
| $\Omega$ | angular velocity of Earth's rotation. |
| $\Omega_i$ | Earth's rotation vector (parallel to the polar axis) |

## Basic empirical dimensionless constants of the EFB closure

| | |
|---|---|
| $C_0 = 0.125$ | inter-component energy exchange constant determining vertical share of TKE, Eqs. (49), (50c) |
| $C_1 = 0.25$, $C_2 = 1.01$ | inter-component energy exchange constants determining longitudinal and transverse shares of TKE, Eqs. (48)-(50) |
| $C_F = 0.125$ | dissipation time-scale constant for turbulent flux of potential temperature, Eq. (19) |
| $C_P = 0.417$ | dissipation time-scale constant for TPE, Eq. (19) |
| $C_r = 1.5$ | standard inter-component energy exchange constant, Eqs. (27), (50a,b,c) |
| $C_\tau = 0.1$ | dissipation time-scale constant for the turbulent flux of momentum, Eq. (33) |
| $C_\Omega = 1$ | rotational length-scale constant, Eq. (73) |
| $R_\infty = 0.25$ | upper limit for the flux Richardson number attainable in the steady-state regime of turbulence, Eqs. (40), (46) |
| $k = 0.4$ | von Karman constant, Eq. (68) |

## Additional constants expressed through the basic constants

| | |
|---|---|
| $a_1 = 0.18$, $a_2 = 0.16$, $a_3 = 1.42$ in Eqs. (81)-(86) | |
| $C_u = k / R_\infty = 1.6$ | in the velocity gradient formulation, Eq. (70) |

$C_\theta = 0.216$                in Eqs. (36), (37), (47), (64)

$k_T = (C_F/C_\tau)k = 0.5$    von Karman constant for temperature, Eq. (86)

$Pr_T^{(0)} = 0.8$             turbulent Prandtl number in neutral stratification, Eq. (57)

$\Pi_\infty = 0.14$             upper limit for the energy stratification parameter, Eq. (77)

## Abbreviations

DNS             direct numerical simulation
EFB             energy- and flux-budget
LES             large-eddy simulation
TKE             turbulent kinetic energy
TPE             turbulent potential energy
TTE             turbulent total energy

## 1. Introduction

Historical overviews of the turbulence closure problem and recent developments in this area of knowledge have been discussed during the last decade by Canuto (2002, 2009), Canuto et al., (2001, 2005, 2008), Cheng et al. (2002), Kurbatskii and Kurbatskaya (2006, 2009, 2010) and Zilitinkevich et al. (2007, 2008, 2009). Most of the operationally used closures employ the concept of the downgradient turbulent transport, implying that the vertical turbulent fluxes of momentum $\tau_i$ ($i=1,2$), potential temperature $F_z$ and other scalars are proportional to their mean gradients. The proportionality coefficients in such relations, called eddy viscosity $K_M$, eddy conductivity $K_H$ and eddy diffusivity $K_D$, are just the unknowns to be determined from the turbulence closure theory. The modern content of this theory originates from Kolmogorov (1941, 1942). He employed the budget equation for the turbulent kinetic energy per unit mass (TKE) $E_K$ to quantify the intensity of turbulence, and postulated that the turbulent exchange coefficients $K_M$, $K_H$ and $K_D$ are fully characterised by the turbulent velocity scale $u_T$, defined as the square root of TKE $u_T = E_K^{1/2}$, and the turbulent time scale $t_T$, defined as the ratio $t_T = E_K/\varepsilon_K$ (where $\varepsilon_K$ is the TKE dissipation rate). This concept has yielded the relations:

$$\varepsilon_K = \frac{E_K}{t_T}, \tag{1}$$

$$K_M \sim K_H \sim K_D \sim u_T^2 t_T \sim u_T l, \tag{2}$$

where $l = E_K^{1/2} t_T$ is the turbulent length scale, whereas the omitted proportionality coefficients in Eq. (2) are assumed to be universal dimensionless constants.

This approach, although quite successful as applied to the neutrally stratified flows, is not quite applicable to the stable stratification. Indeed, Eq. (2) implies that the turbulent Prandtl number $Pr_T \equiv K_M / K_H$ is nothing but universal constant. In the context of the Kolmogorov type closures based on the sole use of the TKE budget equation, this inevitably implies the total decay of turbulence already at moderately stable stratification. However, numerous experiments, large-eddy simulations (LES) and direct numerical simulations (DNS) demonstrate that $Pr_T$ drastically increases with increasing static stability (see Figure 5 below) and, moreover, that turbulence is continuously maintained by the velocity shear even in very stable stratification. This contradiction was overtaken heuristically, prescribing essentially different stability dependences of the turbulent length scales for momentum $l_M$ and heat $l_H$ (and, therefore, for the time scales $t_M$ and $t_H$). In so doing, the Kolmogorov turbulence closure, originally formulated and justified for the neutrally stratified boundary-layer flows (where $l$ can be taken proportional to the distance, *z*, over the surface) factually became unclosed.

In the EFB closure (Zilitinkevich et al., 2007, 2008, 2009) we refined budget equations for the basic second moments: the two energies, the TKE $E_K$ and the turbulent potential energy (TPE) $E_P$, and the vertical turbulent fluxes of momentum and potential temperature, $\tau_i$ ($i = 1,2$) and $F_z$; got rid of the artificial turbulence cut-off in the "supercritical" stratification (inherent to the "one energy equation approach"); and, instead of the traditional postulation of the down-gradient turbulent transport, derived the flux-profile relationships and determined the eddy viscosity and eddy conductivity from the steady-state version of the budget equations for $\tau_i$ and $F_z$.

In the present paper we further advance the physical background of the EFB closure, introduce a new prognostic equation for the turbulent dissipation time scale $t_T$, and extend the theory to non-steady turbulence regimes accounting for non-gradient and non-local turbulent transports (when the traditional concepts of eddy viscosity and eddy conductivity become generally inconsistent).

In Section 2, we refine our approximation of the basic energy- and flux-budget equations, in particular, accounting for the difference between the dissipation time scales for TKE and TPE. In Section 3, focused on the steady-state (algebraic) version of the closure, we develop a new model of the inter-component exchange of TKE (instead of the traditional hypothesis of "return to isotropy" shown to be inconsistent with modern experimental evidence); clarify the concept of the turbulent dissipation time-scale and determine its stability dependence; demonstrate how the steady-state version of the EFB closure relates to the Monin-Obukhov (1954) similarity theory; verify the EFB closure against available empirical data and determine dimensionless universal constants of the theory. In Section 4, we extend the theory to non-steady regimes of turbulence; introduce a relaxation equation for the turbulent dissipation time scale; and propose a hierarchy of the EFB closure models including its most general version based on prognostic equations for all essential parameters of turbulence: $E_K$, $E_P$, $\tau_i$, $F_z$ and $t_T$, and simpler versions employing the concepts of eddy viscosity and eddy conductivity.

We recall that the background stratification of density is characterised by the gradient Richardson number:

$$Ri \equiv \frac{N^2}{S^2}, \tag{3}$$

where $S$ and $N$ are the velocity shear and the Brunt-Väisälä frequency:

$$S^2 = \left(\frac{\partial U}{\partial z}\right)^2 + \left(\frac{\partial V}{\partial z}\right)^2, \tag{4}$$

$$N^2 = \frac{g}{\rho_0}\frac{\partial \rho}{\partial z} = \beta \frac{\partial \Theta}{\partial z}. \tag{5}$$

Here, $z$ is the height; $U$ and $V$ are the mean velocity components along the horizontal axes $x$ and $y$; $\rho$ is the mean density; $\rho_0$ is its reference value; $g = 9.81$ m s$^{-1}$ is the acceleration due to gravity; $\beta = g/T_0$ is the buoyancy parameter; $\Theta$ is the mean potential temperature linked to the absolute temperature $T$ by the relation: $\Theta = T(P_0/P)^{1-1/\gamma}$, where $P$ is the pressure, $P_0$ and $T_0$ are reference values of $P$ and $T$, and $\gamma = c_p/c_v = 1.41$ is the ratio of specific heats. In dry air $\rho = \beta\Theta$, so that the density stratification is fully controlled by the vertical gradient of potential temperature.

Since Richardson (1920), it was generally believed that in stationary homogeneous flows the velocity shear becomes incapable of maintaining turbulence (which therefore collapses) when $Ri$ exceeds some critical value, $Ri_c$ (with conventional value of $Ri_c = 0.25$). On the contrary, in atmospheric and ocean modelling, the turbulence cut-off at "supercritical" values of $Ri$ was understood as an obvious artefact and prevented with the aid of "correction coefficients" specifying the ratios $K_M/(u_T l)$ and $K_H/(u_T l)$ as essentially different functions of $Ri$ (Mellor and Yamada, 1974). The EFB closure automatically accounts for the maintenance of turbulence by the velocity shear at any $Ri$ and does not require any artificial tricks to prevent the turbulence cut-off. It does not imply any critical $Ri$ in the traditional sense (as the boundary between turbulent and laminar regimes) but discloses, just around $Ri \sim 0.2$-$0.3$, quite sharp transition between the two turbulent regimes of principally different nature: strong turbulence at small $Ri$ and weak turbulence at large $Ri$. Following the EFB closure (Elperin et al., 2005; Zilitinkevich et al., 2007), other recently published turbulent closure models (Mauritsen et al. 2007, Canuto et al., 2008, L'vov et al. 2008, Sukoriansky and Galperin, 2008) also do not imply critical Richardson numbers.

## 2. Basic equations

### 2.1. Geophysical approximation

Below we formulate the EFB closure in terms of atmospheric flows characterised by the following typical features:
- Vertical scales of motions (maximum ~10 kilometres) are much smaller than horizontal scales (up to dozens thousands kilometres), which is why the mean-flow vertical velocity, $W$, is orders of magnitude smaller than the horizontal velocities, $U$ and $V$. Hence the vertical turbulent transports are comparable with or even dominate the mean flow vertical advection, whereas the stream-wise horizontal turbulent transport is usually negligible compared to the horizontal advection.
- Typical vertical gradients (along $x_3$ or $z$ axis) of the mean wind velocity $\mathbf{U} = (U_1, U_2, U_3) = (U, V, W)$, potential temperature $\Theta$ and other variables are orders of magnitude larger than the horizontal gradients (along $x_1, x_2$ or $x, y$ axes). Hence, direct effects of the mean-flow horizontal gradients on turbulent statistics are negligible; and the TKE generation is controlled almost entirely by the two components of the velocity gradient: $\partial U / \partial z$ and $\partial V / \partial z$.

Therefore only the components $\tau_{13} = \langle uw \rangle$, $\tau_{23} = \langle vw \rangle$ of the Reynolds stresses $\tau_{ij} = \langle u_i u_j \rangle$ and the vertical component $F_3 = F_z = \langle \theta w \rangle$ of the potential temperature flux $F_i = \langle \theta u_i \rangle$ are needed to close the Reynolds-averaged momentum equations:

$$\frac{DU}{Dt} = fV - \frac{1}{\rho_0}\frac{\partial P}{\partial x} - \frac{\partial \tau_{13}}{\partial z}, \tag{6}$$

$$\frac{DV}{Dt} = -fU - \frac{1}{\rho_0}\frac{\partial P}{\partial y} - \frac{\partial \tau_{23}}{\partial z}, \tag{7}$$

and the thermodynamic energy equation:

$$\frac{D\Theta}{Dt} = -\frac{\partial F_z}{\partial z} + J. \tag{8}$$

Here, $D/Dt = \partial/\partial t + U_k \partial/\partial x_k$, $t$ is the time, $f = 2\Omega \sin\varphi$ is the Coriolis parameter, $\Omega_i$ is the Earth's rotation vector parallel to the polar axis ($|\Omega_i| \equiv \Omega = 0.76 \cdot 10^{-4}$ s$^{-1}$), $\varphi$ is the latitude, $\rho_0$ is the mean density, $J$ is the heating/cooling rate ($J = 0$ in adiabatic processes), $P$ is the mean pressure, $\mathbf{u} = (u_1, u_2, u_3) = (u, v, w)$ and $\theta$ are the velocity and the potential-temperature fluctuations; and angle brackets denote the ensemble-averaged values [see e.g. Holton (2004), Kraus and Businger (1994)]. Generally, atmospheric dynamics problems include the specific-humidity equation [analogous to Eq. (8)], which involves the vertical turbulent flux of humidity $F_q$ contributing to the vertical turbulent flux of buoyancy:

$F_z\beta + 0.61gF_q$. As concerns the turbulence closure, this does not cause additional problems.

General forms of the budget equations for the Reynolds stress, potential-temperature flux and the "energy" of the potential temperature fluctuations $E_\theta = \langle\theta^2\rangle/2$ are

$$\frac{D\tau_{ij}}{Dt} + \frac{\partial}{\partial x_k}\Phi_{ijk}^{(\tau)} = -\tau_{ik}\frac{\partial U_j}{\partial x_k} - \tau_{jk}\frac{\partial U_i}{\partial x_k} - \left[\varepsilon_{ij}^{(\tau)} - \beta(F_j\delta_{i3} + F_i\delta_{j3}) - Q_{ij}\right], \tag{9}$$

$$\frac{DF_i}{Dt} + \frac{\partial}{\partial x_j}\Phi_{ij}^{(F)} = \beta\delta_{i3}\langle\theta^2\rangle - \frac{1}{\rho_0}\langle\theta\nabla_i p\rangle - \tau_{ij}\frac{\partial\Theta}{\partial z}\delta_{j3} - F_j\frac{\partial U_i}{\partial x_j} - \varepsilon_i^{(F)}, \tag{10}$$

$$\frac{DE_\theta}{Dt} + \nabla\cdot\mathbf{\Phi}_\theta = -F_z\frac{\partial\Theta}{\partial z} - \varepsilon_\theta, \tag{11}$$

where $\delta_{ij}$ is the unit tensor ($\delta_{ij} = 1$ for $i = j$ and $\delta_{ij} = 0$ for $i \neq j$); see, e.g., Kaimal and Fennigan (1994), Kurbatsky (2000), Cheng et al. (2002). Other notations in Eqs. (9)-(11) are as follows:

$\Phi_{ijk}^{(\tau)}$, $\Phi_{ij}^F$ and $\Phi_\theta$ are the third-order moments describing turbulent transports of the second-order moments:

$$\Phi_{ijk}^{(\tau)} = \langle u_i u_j u_k\rangle + \frac{1}{\rho_0}\left(\langle pu_i\rangle\delta_{jk} + \langle pu_j\rangle\delta_{ik}\right), \tag{12}$$

$$\Phi_{ij}^{(F)} = \langle u_i u_j \theta\rangle, \tag{13}$$

$$\mathbf{\Phi}_\theta = \frac{1}{2}\langle\theta^2 \mathbf{u}\rangle; \tag{14}$$

$Q_{ij}$ are correlations between fluctuations of the pressure, $p$, and the velocity shear, $\partial u_i/\partial x_j$:

$$Q_{ij} = \frac{1}{\rho_0}\left\langle p\left(\frac{\partial u_i}{\partial x_j} + \frac{\partial u_j}{\partial x_i}\right)\right\rangle; \tag{15}$$

$\varepsilon_{ij}^{(\tau)}$, $\varepsilon_i^{(F)}$ and $\varepsilon_\theta$ are the terms associated with the kinematic viscosity $\nu$ and the temperature conductivity $\kappa$:

$$\varepsilon_{ij}^{(\tau)} = 2\nu\left\langle\frac{\partial u_i}{\partial x_k}\frac{\partial u_j}{\partial x_k}\right\rangle, \tag{16}$$

$$\varepsilon_i^{(F)} = -\kappa \left( \langle u_i \, \Delta \theta \rangle + \Pr \langle \theta \, \Delta u_i \rangle \right), \tag{17}$$

$$\varepsilon_\theta = -\kappa \langle \theta \, \Delta \theta \rangle, \tag{18}$$

where $\Pr = \nu/\kappa$ is the Prandtl number.

The terms $\varepsilon_{ii}^{(\tau)}$, $\varepsilon_i^{(F)}$ and $\varepsilon_\theta$ are essentially positive and represent the dissipation rates of the statistical moments under consideration. Following Kolmogorov (1941, 1942), they are taken proportional to the moments in question divided by the dissipation time scale $t_T$:

$$\varepsilon_{ii}^{(\tau)} = \frac{\tau_{ii}}{2t_T}, \quad \varepsilon_i^{(F)} = \frac{F_i}{C_F \, t_T}, \quad \varepsilon_\theta = \frac{E_\theta}{C_P \, t_T}, \tag{19}$$

where $\tau_{ii} \equiv \langle u_i^2 \rangle$, $C_P$ and $C_F$ are dimensionless universal constants quantifying the difference between the dissipation time scales for different moments.

## 2. EFB model equations

From this point onwards we limit our analysis to the geophysical approximation and basically follow our prior papers (Zilitinkevich et al., 2007, 2008, 2009). The diagonal terms of the Reynolds stress tensor $\tau_{ii} \equiv \langle u_i^2 \rangle$ make doubled components of TKE: $E_i \equiv \langle u_i^2 \rangle / 2$. Their budgets are expressed by Eq. (9) for $i = j$:

$$\frac{DE_i}{Dt} + \frac{\partial}{\partial z} \Phi_i = -\tau_{i3} \frac{\partial U_i}{\partial z} + \frac{1}{2} Q_{ii} - \frac{E_K}{3t_T} \quad (i = 1,2), \tag{20}$$

$$\frac{DE_z}{Dt} + \frac{\partial}{\partial z} \Phi_z = \beta F_z + \frac{1}{2} Q_{33} - \frac{E_K}{3t_T}, \tag{21}$$

where

$$\Phi_i = \frac{1}{2} \langle u_i^2 w \rangle \quad (i = 1,2), \tag{22}$$

$$\Phi_3 = \frac{1}{2} \langle w^3 \rangle + \frac{1}{\rho_0} \langle pw \rangle. \tag{23}$$

Summing up Eqs. (20) and (21), yields the familiar TKE budget equation:

$$\frac{DE_K}{Dt} + \frac{\partial}{\partial z}\Phi_K = -\tau_{i3}\frac{\partial U_i}{\partial z} + \beta F_z - \frac{E_K}{t_T}, \tag{24}$$

where the third term on the right hand side ( r.h.s.) represents the TKE dissipation rate:

$$\varepsilon_K = \frac{E_K}{t_T}, \tag{25}$$

and $\Phi_K$ represents the vertical turbulent flux of TKE:

$$\Phi_K = \frac{1}{2}\langle u_i u_i w\rangle + \frac{1}{\rho_0}\langle p\, w\rangle. \tag{26}$$

The sum of the terms $Q_{ii}$ (the trace of the tensor $Q_{ij}$) vanishes because of the continuity equation: $\partial u_i / \partial x_i = 0$. Hence, these terms are neither productive nor dissipative and describe the kinetic energy exchange between the "richer" component (fed by shear) and the "poorer" transverse and vertical components. Traditionally they were determined through the "return-to-isotropy" hypothesis (Rotta, 1951):

$$Q_{ii} = -\frac{2(C_r + 1)}{3t_T}(3E_i - E_K), \tag{27}$$

where the coefficient $C_r$ was treated as universal dimensionless constant accounting for the difference between the energy-transfer and the energy-dissipation time scales. This formulation contradicts modern experimental evidence even in neutral stratification, where it erroneously prescribes equal shares of the transverse and vertical velocity fluctuations. Moreover, it implies that the share of the transverse velocity fluctuations $E_y/E_K$ does not depend on *Ri*, whereas in reality it significantly increases with increasing *Ri* so that $E_y/E_K$ and $E_x/E_K$ gradually approach each other (see Figure 3 below). In the present paper we develop a new energy exchange concept accounting for these effects.

Although the budget equation for the squared fluctuation of potential temperature $E_\theta$, Eq. (11), was known over decades [see Lumley and Panofsky (1964), Tennekes and Lumley (1972)], its crucial importance for the turbulence energetics was long overlooked. Ostrovsky and Troitskaya (1987) and more recently Zilitinkevich et al. (2007) emphasised close relation between $E_\theta$ and turbulent potential energy. On the background of stable stratification characterised by the Brunt-Väisälä frequency *N*, the vertical displacement of a fluid parcel from its initial level $z$ to the level $z+\delta z$ causes the density increment $\delta\rho = (\partial\rho/\partial z)\delta z = (\rho_0/g)N^2\delta z$, where $\rho$ is the mean density. Then the increment in potential energy per unit mass $\delta E_P = (1/\delta z)\int_z^{z+\delta z}(g/\rho_0)\delta\rho\, z\, dz$ is expressed as $\delta E_P =$

$\frac{1}{2}[(g/\rho_0)\delta\rho]^2/N^2 = \frac{1}{2}(\beta\delta\theta)^2/N^2 = (\beta/N)^2\delta E_\theta$, where $\delta E_\theta = \frac{1}{2}(\delta\theta)^2$ is the increment in the "energy" of the potential temperature fluctuations. This yields the expression of the turbulent potential energy (TPE):

$$E_P = \left(\frac{\beta}{N}\right)^2 E_\theta. \tag{28}$$

In contrast to the potential energy of the mean flow, which depends on the temperature variation linearly, the TPE is proportional to the squared temperature fluctuation. This reminds the concept of available potential energy determined by Lorenz (1955) as the part of the total potential energy of the general circulation available for conversion into kinetic energy. The same is true for the TPE: it is just the potential energy that can be converted into TKE and vice versa.

In geophysical approximation, the budget equation for $E_\theta$, Eq. (11), and the corresponding equation for $E_P$ read:

$$\frac{DE_\theta}{Dt} + \frac{\partial}{\partial z}\Phi_\theta = -F_z \frac{\partial \Theta}{\partial z} - \frac{E_\theta}{C_P t_T}, \tag{29}$$

$$\frac{DE_P}{Dt} + \frac{\partial}{\partial z}\Phi_P = -\beta F_z - \frac{E_P}{C_P t_T}, \tag{30}$$

where $\Phi_\theta$ and $\Phi_P$ are the third-order turbulent fluxes of the second-order fluxes $E_\theta$ and $E_P$, respectively:

$$\Phi_P = \left(\frac{\beta}{N}\right)^2 \Phi_\theta = \frac{1}{2}\left(\frac{\beta}{N}\right)^2 \langle \theta^2 w \rangle. \tag{31}$$

The last terms on the r.h.s. of Eqs. (29) and (30) are the dissipation rates: $\varepsilon_\theta = E_\theta/(C_P t_T)$ and $\varepsilon_P = E_P/(C_P t_T)$.

The buoyancy flux, $\beta F_z$, appears in Eqs. (24) and (30) with opposite signs and describes nothing but the energy exchange between TKE and TPE. In the budget equation for the total turbulent energy (TTE = TKE + TPE), defined as

$$E = E_K + E_P = \frac{1}{2}\left(\langle u_i u_i \rangle + \left(\frac{\beta}{N}\right)^2 \langle \theta^2 \rangle\right), \tag{32}$$

the terms $\pm\beta F_z$ cancel each other. Thus there are no grounds to consider the buoyancy-flux term in the TKE equation as an ultimate "killer" of turbulence.

In Eq. (9) for the vertical components of the turbulent flux of momentum, $\tau_{i3}$ ($i = 1,2$), the molecular-viscosity term, $\varepsilon_{i3}^{(\tau)}$, is small [because the smallest eddies associated with viscous dissipation are presumably isotropic; see L'vov et al. (2009)]; and the dissipative role is played by the combination of terms $\varepsilon_{i3(\text{eff})}^{(\tau)} = -\beta F_i - Q_{i3}$. Zilitinkevich et al. (2007) called this combination the "effective dissipation rate" and expressed it through the Kolmogorov closure hypothesis:

$$\varepsilon_{i3(\text{eff})}^{(\tau)} \equiv -\beta F_i - Q_{i3} = \frac{\tau_{i3}}{C_\tau t_T} = \frac{\tau_{i3} \varepsilon_K}{C_\tau E_K}, \tag{33}$$

where $C_\tau$ is the effective-dissipation time-scale constant. Then the budget equation for $\tau_{i3}$ ($i = 1,2$) simplifies to

$$\frac{D\tau_{i3}}{Dt} + \frac{\partial}{\partial z}\Phi_i^{(\tau)} = -2E_z \frac{\partial U_i}{\partial z} - \frac{\tau_{i3}}{C_\tau t_T}, \tag{34}$$

where $\Phi_i^{(\tau)}$ is the vertical turbulent flux of $\tau_{i3}$:

$$\Phi_i^{(\tau)} = \langle u_i w^2 \rangle + \frac{1}{\rho_0} \langle p\, u_i \rangle. \tag{35}$$

In Zilitinkevich et al. (2007), the concept of the effective dissipation, Eq. (33), was based on our prior analysis of the Reynolds stresses equation in $k$-space using familiar "$\tau$-approximation" (Elperin et al., 2002, 2006). In Figure 1 we compare Eq. (33) with data from large-eddy simulation (LES) of the two types of atmospheric boundary layers: "nocturnal stable" (NS, with essentially negative buoyancy flux at the surface and the neutral stratification in the free flow) and "conventionally neutral" (CN, with negligibly small buoyancy flux at the surface and essentially stable static stability in the free flow). Admittedly, LES is unable to directly reproducing $\varepsilon_K$, which is why we estimated the r.h.s. of Eq. (33) approximately, taking $\varepsilon_K = -\tau_{i3}\partial U_i/\partial z + \beta F_z$ – as it follows from the steady-state version of Eq. (24). In spite of quite large spread of data points, Figure 1 confirms that the effective dissipation $\varepsilon_{13(\text{eff}),\text{definition}}^{(\tau)} \equiv -\beta F_1 - Q_{13}$ (abscissa) is basically proportional to the combination $\tau_{13}\varepsilon_K / E_K$ (ordinate). The grey corridor covering most of data points corresponds to Eq. (33) with $0.1 < C_\tau < 1$, which is consistent with our independent estimate of $C_\tau = 0.2$.

As demonstrated through scaling analysis in Appendix A of Zilitinkevich et al. (2007), the term $\rho_0^{-1}\langle \theta \partial p/\partial z \rangle$ in Eq. (10) for the vertical turbulent flux of potential temperature $F_z$ can be taken proportional to the mean squared temperature, so that

$$\frac{\rho_0^{-1}\langle\theta\,\partial p/\partial z\rangle}{\beta\langle\theta^2\rangle} = 1 - C_\theta, \qquad (36)$$

where $C_\theta$ = constant < 1. In Figure 2 we compare this hypothetical relation with data from LES. Most of the data points (grey corridor) confirm Eq. (36). Then Eq. (10) simplifies to

$$\frac{DF_z}{Dt} + \frac{\partial}{\partial z}\Phi_z^{(F)} = -2(E_z - C_\theta E_P)\frac{\partial\Theta}{\partial z} - \frac{F_z}{C_F\,t_T}. \qquad (37)$$

Equations (20), (21), (24) determine the turbulent kinetic energies $E_i$ ($i = 1,2,3$) and $E_K$; Eqs. (29), (30) determine the "energy" of the temperature fluctuations $E_\theta$ and the turbulent potential energy $E_P$; Eqs. (34), (37) determine the vertical turbulent fluxes of momentum $\tau_{i3}$ ($i = 1,2$) and potential temperature $F_z$. More specifically, the vertical TKE $E_z$ is determined in Section 3.2. The turbulent dissipation time scale $t_T$ is determined in Section 3.4, and the prognostic equation for $t_T$ closing the above system is proposed in Section 4.1.

## 3. Steady-state regime of turbulence

### 3.1. Stability parameters, eddy viscosity and eddy conductivity

We consider the EFB model in its simplest, algebraic form, neglecting non-steady terms in all budget equations. In the TKE budget Equation (24) the first term on the r.h.s. is the rate of the TKE production:

$$-\tau_{i3}\frac{\partial U_i}{\partial z} = \tau S, \qquad (38)$$

where $\tau$ and $S$ are absolute values of the vectors $\boldsymbol{\tau} = (\tau_{xz}, \tau_{yz})$ and $\mathbf{S} = (\partial U/\partial z, \partial V/\partial z)$; and the second term $\beta F_z$ is the rate of conversion of TKE into TPE. The ratio of these terms, called the "flux Richardson number":

$$Ri_f \equiv \frac{-\beta F_z}{\tau S}, \qquad (39)$$

characterises the effect of stratification on turbulence on equal terms with the gradient Richardson number $Ri$, Eq. (3). Clearly, $Ri_f$ can also be treated as the ratio of the velocity-shear length scale $\tau^{1/2}/S$ to the Obukhov (1946) stratification length scale $L$:

$$Ri_f = \frac{\tau^{1/2}}{SL}, \qquad (40)$$

where $L$ is defined as

$$L = \frac{\tau^{3/2}}{-\beta F_z}. \tag{41}$$

Furthermore, the dimensionless height

$$\varsigma = z/L \tag{42}$$

characterises the effect of stratification on equal terms with $Ri$ or $Ri_f$ (Monin and Obukhov, 1954).

The steady-state versions of the budget equations, Eq. (34) and (37), for the vertical turbulent fluxes $\tau_{i3}$ and $F_z$ for the momentum and the potential temperature yield the flux-gradient relations that can be expressed in terms of the eddy viscosity $K_M$ and eddy conductivity $K_H$:

$$\tau_{i3} = -K_M \frac{\partial U_i}{\partial z}, \quad K_M = 2C_\tau E_z t_T, \tag{43}$$

$$F_z = -K_H \frac{\partial \Theta}{\partial z}, \quad K_H = 2C_F t_T E_z \left(1 - C_\theta \frac{E_P}{E_z}\right). \tag{44}$$

The latter relations yield the following expression of the turbulent Prandtl number:

$$Pr_T \equiv \frac{K_M}{K_H} \equiv \frac{Ri}{Ri_f} = \frac{C_\tau}{C_F}\left(1 - C_\theta \frac{E_P}{E_z}\right)^{-1}. \tag{45}$$

It is clearly seen from the steady-state version of the TKE budget Equation (24) that $Ri_f$ in the steady-state regime can only increase with the increasing $Ri$, but obviously cannot exceed unity. Hence it should tend to a finite asymptotic limit (estimated in Section 3.3 as $R_\infty = 0.25$), which corresponds to the asymptotically linear $Ri$-dependence of $Pr_T$:

$$Ri_f \to R_\infty, \quad Pr_T \to \frac{Ri}{R_\infty} \quad \text{at} \quad Ri \to \infty. \tag{46}$$

Similar reasoning including approximation of $Pr_T \approx Pr_t^{(0)} + Ri/R_\infty$ for $Ri \gg 1$, and the estimate of $R_\infty \approx 0.25$ have already been proposed by Schumann and Gerz (1995).

Because $Pr_T \to \infty$ at $Ri \to \infty$, it follows from Eq. (45) that the constant $C_\theta$ [Eqs. (36)-(37)] satisfies the relation

$$C_\theta = \left(E_z / E_P\right)_{Ri \to \infty}, \qquad (47)$$

and therefore is expressed through to other EFB-model constants.

As evident from the above analysis, the concepts of eddy viscosity and eddy conductivity are justified only in the steady state, when the left hand sides (l.h.s.) of the flux budget equations, Eqs. (34) and (37), can be neglected.

### 3.2. Inter-component exchange of turbulent kinetic energy

In the geophysical flows under consideration, the mean wind shear generates the energy of longitudinal velocity fluctuations $E_x$, which feeds the transverse $E_y$ and the vertical $E_z$ energy components. The inter-component energy exchange term in the momentum-flux budget equation, Eq. (9), namely $Q_{ij}$ specified by Eq. (15), is traditionally parameterized through the Rotta (1951) "return-to-isotropy" hypothesis, Eq. (27). In combination with the energy budget Equations (20) and (21), it results in expressions of the longitudinal, $A_x = E_x / E_K$, transverse, $A_y = E_y / E_K$, and vertical, $A_z = E_z / E_K$, shares of TKE characterised by the following features: (i) in neutral stratification $A_z = A_y$; (ii) with strengthening stability $A_x$ increases at the expense of $A_z$ (which therefore decreases), while $A_y$ does not depend on stratification.

However, these features are inconsistent with modern experimental evidence. Available atmospheric data demonstrate that (i) in neutral stratification $A_z^{(0)} \equiv A_z\big|_{\varsigma=0}$ is essentially smaller than $A_y^{(0)} \equiv A_y\big|_{\varsigma=0}$; (ii) with strengthening stability $A_y$ increases and $A_x$ decreases, tending towards horizontal isotropy: $A_y \to A_x$; (iii) the vertical energy share, $A_z$, generally decreases with increasing $\varsigma = z/L$, and at $\varsigma > 1$ levels off at a quite small but non-zero limit (see for example Figure 3). It is conceivable that the stable stratification, suppressing the energy of the vertical velocity $E_z$, facilitates the energy exchange between the horizontal velocity energies $E_y$ and $E_x$, and thereby causes a tendency towards isotropy in the horizontal plane. This newly revealed feature call for revision of the traditional concept of "return-to-isotropy".

We characterise the static stability by the normalised flux Richardson number, $Ri_f / R_\infty$, varying from 0 in neutral stratification to 1 in extremely stable stratification, and propose the following model reflecting the above principal features of the TKE redistribution between the velocity components:

$$Q_{11} = -\frac{2}{3t_T}\left[(C_r+1)(3E_x - E_\Leftrightarrow) + C_r\left(C_1 + C_2\frac{Ri_f}{R_\infty}\right)E_\Leftrightarrow\right], \tag{48a}$$

$$Q_{22} = -\frac{2}{3t_T}\left[(C_r+1)(3E_y - E_\Leftrightarrow) - C_r\left(C_1 + C_2\frac{Ri_f}{R_\infty}\right)E_\Leftrightarrow\right], \tag{48b}$$

$$Q_{33} = -\frac{2(C_r+1)}{t_T}\left(E_z - E_K + \frac{2}{3}E_\Leftrightarrow\right), \tag{48c}$$

where $E_\Leftrightarrow$ is the part of TKE participating in the inter-component energy exchange:

$$E_\Leftrightarrow = E_K + \frac{C_r}{(1+C_r)}\frac{Ri_f}{R_\infty}\left(C_0 E_K - (1+C_0)E_z\right). \tag{49}$$

Substituting the energy exchange model, Eqs. (48)-(49), in the steady-state version of the energy-budget equations (20), (21) yields:

$$A_x = \frac{1}{2(1+C_r)(1-Ri_f)}\left[1 + \frac{C_r(1+2Ri_f)}{3(1+C_r)}\left(C_1 + C_2\frac{Ri_f}{R_\infty}\right)\right] \\ + \frac{1-A_z}{2}\left[1 - \frac{C_r}{(1+C_r)}\left(C_1 + C_2\frac{Ri_f}{R_\infty}\right)\right], \tag{50a}$$

$$A_y = -\frac{1}{2(1+C_r)(1-Ri_f)}\left[1 + \frac{C_r(1+2Ri_f)}{3(1+C_r)}\left(C_1 + C_2\frac{Ri_f}{R_\infty}\right)\right] \\ + \frac{1-A_z}{2}\left[1 + \frac{C_r}{(1+C_r)}\left(C_1 + C_2\frac{Ri_f}{R_\infty}\right)\right], \tag{50b}$$

$$A_z = \frac{E_z}{E_K} = \frac{C_r(1-2C_0\,Ri_f/R_\infty)(1-Ri_f) - 3Ri_f}{(1-Ri_f)\{3 + C_r[3 - 2(1+C_0)Ri_f/R_\infty]\}}, \tag{50c}$$

where $C_0$, $C_1$ and $C_2$ are dimensionless empirical constants. The condition $A_x = A_y$ at $Ri_f = R_\infty$ yields an expression for $C_1 + C_2$:

$$C_1 + C_2 = C_r^{-1}(1-R_\infty)^{-1}\left[1 - A_z^{(\infty)} - \frac{1+2R_\infty}{3(1+C_r)(1-R_\infty)}\right]^{-1}, \tag{50d}$$

while the coefficient $C_1$ is determined at $Ri_f \to 0$:

$$C_1 = \frac{(1+C_r)(A_x^{(0)} - A_y^{(0)}) - 1}{A_z^{(0)}[1 - 3(1+C_r)(1 - A_z^{(0)})]}, \tag{50e}$$

where $A_z^{(0)} = A_z\big|_{Ri=0}$, $A_z^{(\infty)} = A_z\big|_{Ri\to\infty}$, and we took into account that $E_K / K_M S^2 t_T = 1 - Ri_f$. Figure 3 shows the energy shares $A_i$, determined by Eqs. (50) and converted into $z/L$ dependences using Eq. (71) (Section 3.4). Fitting theoretical curves, Eqs. (50a) and (50b), to rather scarce data presented in the figure yields tentative estimates of $C_1 = 0.25$ and $C_2 = 1.01$. In our further analyses they are not needed. Of the TKE shares we use only $A_z$, Eq. (50c), to determine $E_z$ in Equations (43) and (44) for the eddy viscosity and eddy conductivity.

According to Eq. (50c), $A_z$ varies between the following limits:

$$A_z\big|_{Ri=0} = A_z^{(0)} = \frac{C_r}{3(1+C_r)}, \tag{51}$$

$$A_z\big|_{Ri\to\infty} = A_z^{(\infty)} = \frac{C_r(1 - 2C_0) - \dfrac{3R_\infty}{1 - R_\infty}}{3 + C_r(1 - 2C_0)}. \tag{52}$$

Empirical constants $C_0$, $C_r$ and $R_\infty$ are determined below.

### 3.3. Stability dependencies of basic parameters of turbulence, and determination of empirical constants

In the steady state Eqs. (20), (21), (24), (29), (30), (32), (34) and (37) reduce to algebraic system of equations governing local balances between the generation and dissipation terms. Although this system is not closed (until the turbulent time scale $t_T$ is determined), it allows us to determine basic dimensionless parameters of turbulence as universal functions of the gradient Richardson number $Ri$, Eq. (3).

Combining Eqs. (24), (30) and Eq. (32) yields the following expressions of the shares of TKE and TPE as universal functions of the flux Richardson number:

$$\frac{E_K}{E} = \frac{1 - Ri_f}{1 - (1 - C_P)Ri_f}, \tag{53}$$

$$\frac{E_P}{E} = \frac{C_P Ri_f}{1 - (1 - C_P)Ri_f}. \tag{54}$$

Then using Eq. (47) to determine $C_\theta$:

$$C_\theta = \frac{(1-R_\infty)A_z^{(\infty)}}{C_P R_\infty}, \qquad (55)$$

and combining Eqs. (45), (53)-(55) we determine the gradient Richardson number $Ri$ and the turbulent Prandtl number $Pr_T$:

$$Ri = Pr_T Ri_f = \frac{C_\tau}{C_F} Ri_f \left(1 - \frac{Ri_f(1-R_\infty)A_z^{(\infty)}}{R_\infty(1-Ri_f)A_z(Ri_f)}\right)^{-1}. \qquad (56)$$

Equations (50c), (56) determine $Ri$ as the universal infinitely increasing function of $Ri_f$ and, thereby, implicitly determine

- $Ri_f$ as universal monotonically increasing function of $Ri$ approaching $R_\infty$ at $Ri \to \infty$;
- and $Pr_T$ as infinitely increasing function of $Ri$ having the asymptote: $Pr_T \to Ri/R_\infty$ at $Ri \to \infty$.

Comparison of these functions with data in Figures 4 and 5 yields quite certain empirical estimate of $R_\infty = 0.25$ [cf. Schumann and Gerz (1995)], implying very strong asymptotic $Ri$-dependence of the turbulent Prandtl number: $Pr_T \approx 4Ri$ at $Ri \gg 1$.

Data for very small $Ri$ in Figures 4 and 5 are consistent with the well-established empirical value of the turbulent Prandtl number in neutral stratification [e.g., Elperin et al. (1996), Churchill (2002), Foken (2006)]:

$$Pr_T \to Pr_T^{(0)} = \frac{C_\tau}{C_F} = 0.8 \quad \text{at} \quad Ri \to 0. \qquad (57)$$

As follows from Eq. (56) in linear approximation with respect to $Ri$, the turbulent Prandtl number at $Ri \ll 1$ behaves as

$$Pr_T \approx Pr_T^{(0)} + \frac{(1-R_\infty)A_z^{(\infty)}}{R_\infty A_z^{(0)}} Ri. \qquad (58)$$

Taking empirical values of $R_\infty = 0.25$, $A_z^{(0)} = 0.2$ and $A_z^{(\infty)} = 0.03$ (see Figure 6 below), Eq. (58) yields $Pr_T \approx 0.8 + 0.45Ri$. This means that $Pr_T$ in the strong-turbulence regime typical of boundary-layer flows varies insignificantly, increasing from 0.8 at $Ri = 0$ to 0.9 at $Ri = 0.25$. On the background of quite natural spread of data, it is practically impossible to recognise such weak dependence empirically. Over decades, this inherent feature of the boundary-layer turbulence has served as a basis for the widely used assumption $Pr_T =$ constant, which has given the name "Reynolds analogy". Our theory justifies it as a reasonable approximation for the strong-turbulence regime ($0 < Ri < 0.25$), and reveals its absolute inapplicability to the weak-turbulence regime ($Ri > 1$), where the $Ri$-dependence of $Pr_T$ becomes an order of magnitude stronger: $dPr_T/dRi \approx 4$. Zilitinkevich (2010) has already pointed out strongly different $Ri$-dependences of $Pr_T$ at large and small $Ri$ in connection with conceptual inadequacy of the currently used design of DNS of the stably stratified turbulence for small $Ri$.

Owing to Eq. (56), the above Eqs. (50c), (53) and (54) determine the vertical share of TKE $A_z$, and the ratios $E_K/E$ and $E_P/E$ as universal functions of $Ri$. Figure 6 shows empirical data on $A_z$ together with theoretical curve plotted after Eq. (50c). Inspection of this figure yields rough estimates of $A_z^{(0)} = 0.2$ and $A_z^{(\infty)} = 0.03$. Consequently, Eq. (51) yields $C_r = 1.5$; and using the above estimate of $R_\infty = 0.25$, Eq. (52) yields $C_0 = 0.125$.

Figure 7 shows empirical verification of the $Ri$-dependence of $E_P/E$ after Eq. (54). At $Ri \to \infty$ it has the limit:

$$\frac{E_P}{E} \to \frac{C_P R_\infty}{1-(1-C_P)R_\infty}. \tag{59}$$

Empirical data in Figure 7 are basically consistent with the curve and allow for estimating the limit: $E_P/E|_{Ri\to\infty} \to 0.122$. Using the above estimate of $R_\infty = 0.25$, this yields $C_P = 0.417$. We recall that $C_P$ is the ratio of the dissipation time scales for TKE and TPE. Venayagamoorthy and Stretch (2006, 2010) investigated these scales using experimental data on the grid-generated turbulence (Srivat and Warhaft, 1983; Itsweire at al., 1986; Yoon and Warhaft, 1990; Mydlarski, 2003) and data from DNS of the stably stratified (Shih et al., 2000) and neutrally stratified (Rogers at al., 1989) homogeneous sheared turbulence. Their analysis demonstrated that time-scale ratio is relatively insensitive to $Ri$, which supports our treatment of $C_P$ as universal constant.

The steady-state version of Eq. (24) together with Eq. (43) yield the following $Ri_f$-dependence of the dimensionless turbulent flux of momentum:

$$\left(\frac{\tau}{E_K}\right)^2 = \frac{2C_\tau A_z(Ri_f)}{(1-Ri_f)}. \tag{60}$$

The steady-state version of Eq. (29) together with Eqs. (44)-(45) yield the $Ri_f$-dependence of the dimensionless turbulent flux of potential temperature:

$$\frac{F_z^2}{E_K E_\theta} = \frac{2C_\tau}{C_P} \frac{A_z(Ri_f)}{Pr_T}. \tag{61}$$

Here, the function $A_z(Ri_f)$ is determined by Eq. (50c) and the function $Ri_f(Ri)$, by Eq. (56); hence Eq. (60) specifies $Ri$-dependence of $(\tau/E_K)^2$ and Eq. (61), $Ri$-dependence of $F_z^2/(E_K E_\theta)$.

Available data on $(\tau/E_K)^2$ together with the theoretical curve plotted after Eq. (60) are shown in Figure 8. They are consistent with the commonly accepted estimate of $(\tau/E_K)_{Ri \to 0} = 0.2$ (e.g., Monin and Yaglom, 1971) and, in spite of a wide spread, confirm a pronounced decrease in $\tau/E_K$ with increasing $Ri$. Using this figure we roughly estimate $2C_\tau A_z^{(0)} = 0.04$ and (using the above empirical value of $A_z^{(0)} = 0.2$) determine $C_\tau = 0.1$.

Empirical verification of Eq. (61) shown in Figure 9 demonstrates a reasonably good correspondence between the theory and data, and allows for determining the small-$Ri$ limit:

$$\left(\frac{F_z^2}{E_K E_\theta}\right)_{Ri \to 0} = \frac{2C_\tau}{C_P} \frac{A_z^{(0)}}{Pr_T^{(0)}} = \frac{2A_z^{(0)} C_F}{C_P} = 0.12, \tag{62}$$

which yields $C_F/C_P = 0.3$. Since $C_P = 0.417$ is already determined, we get $C_F = 0.125$.

The above estimates provide empirical values of our basic dimensionless constants:

$$C_0 = 0.125,\ C_F = 0.125,\ C_P = 0.417,\ C_r = 1.5,\ C_\tau = 0.1,\ R_\infty = 0.25. \tag{63}$$

We admit that the empirical foundation of these estimates is not quite solid. We deliberately selected data sets shown in different figures to avoid biasing clouds of data points. Our excuse is that the algebraic version of the model selected for validation against empirical data is valid only in the stationary homogeneous turbulence, whereas available data sets (except DNS of the stably stratified turbulence for given Richardson numbers) basically correspond to heterogeneous and/or non-stationary turbulence. Our estimation of the empirical constants from quite limited and not fully reliable data sets is to some extent justified by the facts that the constants are interdependent (changing one of them we are forced to change all others), and the number of constants is less than the number of the employed empirical dependencies. This made it possible to determine the entire set of constants searching for the optimal solution to the over-determined set of algebraic relations expressing the unknown constants through the measurable parameters.

As follows from Eq. (47), the constant $C_\theta = \lim(E_z/E_P)|_{Ri \to \infty}$ is not independent. Then the identity $E_z/E_P = A_z(E/E_P - 1)$, Eq. (52) for $A_z^{(\infty)}$, and our empirical estimate of $\lim(E/E_P)|_{Ri \to \infty} = 8$ [resulted form Eq. (59) and Figure 7] yield:

$$C_\theta = \frac{C_r(1-2C_0)(1-R_\infty) - 3R_\infty}{[1+(C_P-1)R_\infty][3+C_r(1-2C_0)]} = 0.216. \tag{64}$$

The above theoretical results are quite unusual considering that the stability dependences of the dimensionless parameters of turbulence, in particular, the $Ri$-dependencies of the flux Richardson number $Ri_f$ and the turbulent Prandtl number $Pr_T$, given by Eq. (56), are determined from an unclosed system of equations, regardless of the particular formulation of the turbulent dissipation time scale $t_T$. The latter is determined in the next section from asymptotic analysis of the velocity shear and TKE budget in strong- and weak-turbulence regimes.

### 3.4. Turbulent dissipation time and length scales

The time scale $t_T$ or the length scale $l$ appear in the Kolmogorov closure for the dissipation rates, Eqs. (1), (19), (25), (33). Until present, determination of these scales remained one of the most uncertain aspects of the turbulence closure problem. The only simple case, when $l$ is easily determined, is the non-rotating neutrally stratified boundary layer flow over a flat surface, where the turbulent length scale is restricted only by the distance from the surface, $z$. Then the "master length scale" $l_0 = l|_{Ri=0}$ can be taken proportional to $z$:

$$l_0 = l|_{Ri=0} = C_l z, \tag{65}$$

where $C_l$ = constant[1]. In the stable stratification, an additional restriction appears due to the balance between the kinetic energy of a fluid parcel and its potential energy acquired at the expense of displacement. Using the Obukhov length scale $L$, Eq. (41), to quantify this restriction, and leaving aside the restriction caused by the Earth's rotation, it stands to reason that the turbulent length scale $l$ in the stably stratified boundary layer close to the surface monotonically increases with increasing height: $l = l_0 \sim z$ at $z \ll L$, whereas far from the surface it levels off: $l \sim L$ at $z \gg L$.

In view of these two limits, the easiest way to determine $l$ that comes to mind is the interpolation of the type $l \sim z/(1 + \text{constant } z/L)$, employing either the Obukhov length scale $L$ or alternative stratification length scales: $E_K^{1/2}/N$, $\varepsilon_K^{1/2}/N^{3/2}$, etc. However, none of such interpolations has led to satisfactory results. The problem is aggravated by the lack of high-quality data on the stability dependence of $t_T$ or $l$. The point is that $t_T \equiv E_K/\varepsilon_K$ or $l \equiv E_K^{1/2} t_T$ are virtual parameters determined through $E_K$ and $\varepsilon_K$, which both are not easily measurable. Therefore hypothetical interpolation formulae for $t_T$ or $l$ are verified indirectly, through the overall performance of the turbulence closure model. This method does not offer a clear understanding which elements of the closure are correct and which are erroneous.

---

[1] Obukhov (1942) has developed a method for determining the master length scale for complex domains.

Instead, Zilitinkevich et al. (2010) have revealed the stability dependence of the turbulent time scale indirectly from the stability dependence of the velocity shear $S$ determined quite accurately in numerous field experiments and LES. For the neutrally stratified boundary layer flow (with $Ri \ll 1$, $z/L \ll 1$), taking $l = C_l z$ and combining the steady-state version of Eq. (24) with Eqs. (51) and (60) yields the familiar wall law:

$$S = \frac{\tau^{1/2}}{kz}, \tag{66}$$

where $k$ is the von-Karman constant expressed through $C_l$ and other dimensionless constants of the EFB closure:

$$k = C_l \left[ \frac{2C_\tau C_r}{3(1+C_r)} \right]^{3/4}. \tag{67}$$

Adopting the conventional empirical value of $k = 0.4$, yields $C_l = 4.47$. Hereafter we include $k$ instead of $C_l$ in the set of basic empirical constants of the EFB closure.

Alternatively, in very stable stratification (at $Ri > 1$, $z/L \gg 1$), Eqs. (39) and (46) yield the following asymptotic expression of the velocity shear:

$$S = \frac{-\beta F_z}{R_\infty \tau} = \frac{\tau^{1/2}}{R_\infty L}. \tag{68}$$

The straightforward interpolation between Eqs. (66) and (68) reads

$$S = \frac{\tau^{1/2}}{kz} \left( 1 + \frac{k}{R_\infty} \frac{z}{L} \right). \tag{69}$$

Clearly, there are no *a priori* grounds to expect that Eq. (69) is valid between the limits Eq. (66) and Eq. (68). But very luckily it happens to be the case: Eq. (69) shows excellent agreement with experimental data for the steady-state non-rotating sheared flows over the *entire range of stratifications from neutral to extremely stable*.

Indeed, the linear $z/L$ dependence of the "velocity $\Phi$-function":

$$\Phi_M \equiv \frac{kz}{\tau^{1/2}} S = 1 + C_u \frac{z}{L}, \tag{70}$$

established by Monin and Obukhov (1954) for the stably stratified atmospheric surface layer (where $Ri$ varies from 0 to 0.25, and $z/L$ varies from 0 to 10), was confirmed in numerous experiments (e.g., Monin and Yaglom, 1971) and LES that yielded quite solid estimates of the empirical constants $k \approx 0.4$, $C_u \approx 1.6$ (see Figure 10). On the other hand, adopting the conventional empirical value of $k \approx 0.4$ and the estimate of $R_\infty \approx 0.25$ based on experimental, LES and DNS data for the very stably stratified flows (covering a wide range of $Ri$ from 1 to $10^2$), the empirical constant $k/R_\infty$ on the r.h.s. of Eq. (69) [precisely analogous to $C_u$ in Eq. (70)] is also estimated as $k/R_\infty \approx 1.6$. What this means is that Eq. (69) agrees very well with experimental data on the velocity gradient over the entire range of stratifications from $Ri < 0.25$ (in the atmospheric surface layer) up to $Ri \sim 10^2$ (in LES, DNS and lab experiments). On these grounds Eq. (69) can be considered as a firmly established feature of the locally balanced steady-state stably stratified sheared flows.

Combining Eq. (69) with the definition of the flux Richardson number, Eq. (40), yields the following relations linking $Ri_f$ and $z/L$:

$$Ri_f = \frac{kz/L}{1+kR_\infty^{-1}z/L}, \quad \frac{z}{L} = \frac{R_\infty}{k} \frac{Ri_f}{R_\infty - Ri_f}. \tag{71}$$

Furthermore, substituting $-\tau_{i3}\partial U_i/\partial z = \tau S$ after Eq. (69) into the steady-state version of the TKE budget equation, Eq. (24), and accounting for Eq. (71), yields the stability dependence of the turbulent dissipation time and length scales, $t_T$ and $l$, in terms of either $z/L$ or $Ri_f$:

$$l = t_T E_K^{1/2} = kz \frac{(E_K/\tau)^{3/2}}{1+k(R_\infty^{-1}-1)z/L} = kz \left(\frac{E_K}{\tau}\right)^{3/2} \frac{1-Ri_f/R_\infty}{1-Ri_f}, \tag{72}$$

where $E_K/\tau$ is expressed by Eq. (60) as universal function of $Ri_f$ [that can be converted into a function of $z/L$ using Eq. (71)]. Equation (72) has quite expectable asymptotes: $l \sim z$ for $z/L \to 0$, and $l \sim L$ for $z/L \to \infty$. However, it essentially differs from the mere linear interpolation between $1/z$ and $1/L$, since the factor $(E_K/\tau)^{3/2}$ on the r.h.s. of Eq. (72) strongly increases with increasing stability and approaches a finite limit only at $Ri > 1$, that is outside geophysical boundary-layer flows, where $Ri$ is typically less than 0.25 (see empirical $Ri$-dependence of $E_K/\tau$ in Figure 8).

Besides the effect of stratification, $l$ and $t_T$ are affected by the angular velocity of Earth's rotation $\Omega = 7.29 \cdot 10^{-5}$ s$^{-1}$, which involves the rotational length-scale limit: $E_K^{1/2}/\Omega$. Accordingly, we determine the master length $l_0$ interpolating between the surface limit, Eq. (65) and the above mentioned rotational limit, which yields $l_0 = C_l z/(1+C_\Omega \Omega z/E_K^{1/2})$, where $C_\Omega$ is empirical dimensionless constant. Then Eq. (72) becomes

$$l = t_T E_K^{1/2} = \frac{kz}{1 + C_\Omega \Omega z / E_K^{1/2}} \frac{(E_K/\tau)^{3/2}}{1 + k(R_\infty^{-1} - 1)z/L}$$

$$= \frac{kz}{1 + C_\Omega \Omega z / E_K^{1/2}} \left(\frac{E_K}{\tau}\right)^{3/2} \frac{1 - Ri_f / R_\infty}{1 - Ri_f}. \quad (73)$$

Blackadar (1962) was probably the first who called attention to the effect of the Earth's rotation on the turbulent length scale. He proposed a relation analogous to Eq. (73) with the only difference that the rotational turbulent-length-scale limit was defined through the ratio $U/f$, where $U$ is the mean wind velocity (rather than turbulent velocity scale $E_K^{1/2}$) and $f = 2\Omega \sin\varphi$ is the Coriolis parameter (rather than the angular velocity of Earth's rotation $\Omega$). In our notations Blackadar's relation becomes $l_0 = C_l z/(1 + C_B fz/U)$, where $C_B$ is empirical dimensionless coefficient. Relying upon its commonly accepted empirical value $C_B = 1.5 \cdot 10^3$ (e.g., Sorbjan, 2011) and accounting for the typical value of the intensity of turbulence in the free atmosphere $E_K^{1/2}/U$ ~ $10^{-3}$, yields a rough estimate of our dimensionless constant: $C_\Omega$ ~ $C_B(E_K^{1/2}/U)$ ~ 1.

We do not strictly follow Blackadar (1962) because $E_K^{1/2}$ is obviously more relevant than $U$ as the turbulent velocity scale; and $\Omega$ is more relevant than $f$ as the rotational frequency scale. Indeed, $f$ characterises exclusively the vertical component of the vector $\Omega_i$ ($i = 1,2,3$), which affects the horizontal velocity components, whereas turbulent motions are essentially 3-dimensional and are effected by all three components of $\Omega_i$ (see Glazunov, 2010).

It is significant that the traditional stratification parameters $Ri_f = -\tau S/F_z$ and $z/L = -\beta F_z z/\tau^{3/2}$, widely used in boundary-layer physics, are based on the local values of turbulent fluxes $\tau$ and $F_z$. In the context of turbulence closure problem, these are just the unknown parameters to be determined. Therefore closure models formulated in terms of $Ri_f$ or $z/L$ imply iteration procedures with no guarantee that errors in determining $\tau$ and $F_z$ (in very stable stratification comparable with $\tau$ and $F_z$ as such) would not disrupt convergence of iterations. To overcome this difficulty, we propose a new *energy stratification parameter*:

$$\Pi = E_P / E_K. \quad (74)$$

The steady-state versions of Eqs. (24) and (30) allow for expressing $Ri_f$ through $\Pi$ and vice versa:

$$Ri_f = \frac{\Pi}{C_P + \Pi}. \quad (75)$$

In terms of $\Pi$, Eq. (73) becomes

$$t_{TE} = \frac{l}{E_K^{1/2}} = \frac{kz}{E_K^{1/2} + C_\Omega \Omega z} \left(\frac{E_K}{\tau}\right)^{3/2} \left(1 - \frac{\Pi}{\Pi_\infty}\right). \tag{76}$$

Here, $\Pi_\infty = C_P R_\infty /(1 - R_\infty) = 0.139$ is the maximal value of $\Pi$ corresponding to extremely stable stratification, and the additional subscript "$E$" in $t_{TE}$ indicates that Eq. (76) determines the turbulent dissipation time scale $t_T$ in the equilibrium state corresponding to local balance between the production and the dissipation rates of turbulence. The ratio $E_K / \tau$ is determined after Eqs. (60) and (75):

$$\left(\frac{E_K}{\tau}\right)^2 = \frac{C_P}{2C_\tau (C_P + \Pi) A_z}; \tag{77}$$

and the vertical share of TKE $A_z$, is determined after Eqs. (50c) and (75):

$$A_z = \frac{E_z}{E_K} = \frac{\Pi_\infty (C_r - 3\Pi/C_P)(C_P + \Pi) - 2C_r C_0 (C_P + \Pi_\infty)\Pi}{3\Pi_\infty (1 + C_r)(C_P + \Pi) - 2C_r (1 + C_0)(C_P + \Pi_\infty)\Pi}. \tag{78}$$

As evidenced by Eqs. (50c) [or (78)], $A_z$ monotonically decreases with increasing stability and at $Ri_f \to R_\infty$ [or $\Pi \to \Pi_\infty$] tends to a finite positive limit; whereas $t_{TE}$, Eq. (72) diminishes to zero. Equations (76)-(78) close the algebraic version of the EFB closure. Clearly, determining $t_{TE}$ (or $l$) is fully equivalent to the determining the TKE dissipation rate $\varepsilon = E_K / t_K$.

### 3.5. Application to boundary-layer turbulence

Equation (71) links the flux Richardson number $Ri_f$ with the dimensionless height $\varsigma = z/L$ based on the Obukhov length scale $L$, Eq. (42). This relation is valid not too far from the surface, namely at $z \ll E_K^{1/2}/\Omega$, where the master length scale $l_0$, Eq. (73), reduces to $C_l z$. However, in the upper part of the atmospheric boundary layer the effect of $\Omega$ on the master length scale $l_0$ can be significant. Indeed, the TKE at the upper boundary of the layer becomes very small compared to its near-surface value. Taking $\Omega = 7.29 \cdot 10^{-5}$ s$^{-1}$ and adopting a rough estimate of $E_K^{1/2}\big|_{z=h} \sim 0.1$ m s$^{-1}$ yields the rotational length scale $E_K^{1/2}/\Omega \sim 10^3$ m, which is quite comparable with the typical boundary-layer height $h \sim 5 \cdot 10^2$ m. Anyhow, close to the surface the effect of rotation on $l_0$ is obviously negligible. Hence, using Eq. (71), the dimensionless parameters of turbulence, presented in Sections 3.2 and 3.3 as universal functions of $Ri_f$, can be reformulated as universal functions of $\varsigma = z/L$.

The concept of similarity of turbulence in terms of the dimensionless height $\varsigma$ has been proposed by Monin and Obukhov (1954) for the "surface layer" defined as the lower one tenths of the boundary layer, where the turbulent fluxes of momentum $\tau$, temperature $F_z$ and other scalars, as well as the length scale $L$, are reasonably accurately approximated by their surface values: $\tau = \tau|_{z=0} \equiv u_*^2$, $F_z = F_z|_{z=0} \equiv F_*$, $L = L|_{z=0} \equiv L_*$. This widely recognised similarity concept was confirmed, particularly for stable stratification, in numerous field and laboratory experiments (see Monin and Yaglom, 1971; Sorbjan, 1989; Garratt, 1992) and more recently through LES and DNS. Nieuwstadt (1984) extended this concept to the entire stable boundary layer employing local $z$-dependent values of the fluxes $\tau$, $F_z$ and the length $L$ instead of their surface values: $u_*^2$, $F_*$ and $L_*$.

The EFB closure as applied to the steady-state non-rotating boundary-layer flows is fully consistent with the Monin-Obukhov and Nieuwstadt similarity theories. Considering the immense available information on the atmospheric boundary-layer turbulence, we present examples of theoretical relationships potentially useful in modelling applications:

the ratio of TPE to TTE, Eq. (54):

$$\frac{E_P}{E} = \frac{C_P R_\infty \varsigma}{R_\infty / k + [1 + (C_P - 1) R_\infty] \varsigma}, \qquad (79)$$

the vertical share of TKE, Eq. (50c):

$$A_z = \frac{E_z}{E_K} = \frac{R_\infty C_r + k\varsigma \left[ C_r (1 - 2C_0) - \dfrac{3R_\infty (R_\infty + k\varsigma)}{R_\infty + k\varsigma (1 - R_\infty)} \right]}{3R_\infty (1 + C_r) + k\varsigma [3 + C_r (1 - 2C_0)]}, \qquad (80)$$

the turbulent Prandtl number, Eq. (56):

$$Pr_T = \frac{C_\tau}{C_F} \left[ 1 + \frac{a_1 \varsigma + a_2 \varsigma^2}{1 + a_3 \varsigma} \right], \qquad (81)$$

and the gradient Richardson number [from Eqs. (71), (81)]:

$$Ri = Ri_f Pr_T = \frac{C_\tau k\varsigma}{C_F (1 + R_\infty^{-1} k\varsigma)} \left[ 1 + \frac{a_1 \varsigma + a_2 \varsigma^2}{1 + a_3 \varsigma} \right], \qquad (82)$$

where $a_1$, $a_2$ and $a_3$ are known empirical constants:

$$a_1 = 3k(1+C_r)\frac{(1-2C_0)(R_\infty^{-1}-1)-3C_r^{-1}}{3+C_r(1-2C_0)}, \tag{83}$$

$$a_2 = \frac{k^2}{R_\infty}\left[(1-2C_0)(R_\infty^{-1}-1)-3C_r^{-1}\right], \tag{84}$$

$$a_3 = \frac{k}{R_\infty}\left(\frac{6(C_0+1)}{3+C_r(1-2C_0)}+2(R_\infty-C_0)-1\right). \tag{85}$$

According to the EFB closure, the mean velocity gradient in the steady-state non-rotating boundary layer flow is expressed by Eq. (69) that implies the following $\varsigma$-dependence of the eddy viscosity: $K_M = \tau/S = k\tau^{1/2}z\left[1+(k/R_\infty)\varsigma\right]^{-1}$. Therefore Eqs. (71) and (85) allow for determining turbulent Prandtl number $Pr_T$, eddy conductivity $K_H = K_M/Pr_T$, potential temperature gradient $\partial\Theta/\partial z = -F_z/K_H$, and the "temperature $\Phi$-function":

$$\Phi_H \equiv \frac{k_T\tau^{1/2}z}{-F_z}\frac{\partial\Theta}{\partial z} = \left[1+\frac{a_1\varsigma+a_2\varsigma^2}{1+a_3\varsigma}\right]\left(1+\frac{k}{R_\infty}\varsigma\right), \tag{86}$$

where $k_T = (C_F/C_\tau)k = 0.5$ is the temperature von-Karman constant, and $k = 0.4$ is the velocity von Karman constant, Eq. (67).

In Figure 11, Eq. (86) is compared with our LES. Because all model constants in Eq. (86) are already determined from other empirical dependencies, very good agreement between the theory and LES data in Figure 11 serves as an independent verification of the EFB model.

Given the velocity and temperature $\Phi$-functions, Eq. (70) and (86), $\varsigma$-dependence of the gradient Richardson number is immediately determined: $Ri = k(C_\tau/C_F)\varsigma\Phi_H/\Phi_M^2$. Its comparison with our LES is shown in Figure 12.

Besides the LES data points, in Figures 10-12 we demonstrate the two versions of the bin averaged data shown as open triangles for $\varsigma = z/L$ and black triangles for $\varsigma = z/[(1+C_\Omega\Omega z/E_K^{1/2})L]$ with $C_\Omega = 1$. In Figures 11 and 12 black triangles obviously better fit theoretical curves at large $\varsigma$. This supports the estimate of $C_\Omega = 1$ and confirms that the Earth rotations starts affecting the turbulent length and time scales already in the upper part of the planetary boundary layer – at a few hundred meter heights.

To the best of our knowledge, none of Eqs. (79)-(86) was obtained before. Moreover, in the traditional interpretation of the Monin-Obukhov similarity theory it was taken as self-evident that the maximal values of $\varsigma = z/L$ achievable in the atmospheric surface layer (factually never exceeding 10) can be attributed to the very strong static stability regime that has given the name "*z*-less stratification regime". Accordingly, it was assumed that at *z* >> *L* the distance over the surface does not affect turbulence and, therefore, should disappear from any similarity-theory relations, for instance, from the expressions on the r.h.s. of Eqs. (79)-(81), (85), which therefore should turn into universal constants.

This reasoning is not quite correct. The point is that the really strongly-stable stratification is principally unattainable neither in the surface layer nor in the atmospheric boundary layers. The boundary-layer flows correspond to quite small gradient Richardson numbers *Ri* << 1 and only moderate dimensionless heights $\varsigma$ << 10. Moreover, even at $\varsigma \to \infty$, the similarity functions do not necessarily turn into finite constants, but can also tend to zero [as for instance the dimensionless heat flux: Eq. (61) and Figure 9] or to infinity [as for instance $\varsigma$-dependencies of *Ri* and $Pr_T$: Eq. (85) and Figure 5]. Factually, atmospheric boundary layers are *weakly-stable* strong-turbulence layers characterised by the gradient Richardson numbers essentially smaller than unity and dimensionless heights $\varsigma = z/L$ smaller than 10. The *strongly-stable* stratification with by *Ri* >> 1 and $\varsigma$ >> $10^2$ corresponds to the weak-turbulence regime typical of the free atmosphere.

## 4. Hierarchy of EFB turbulence closures

### 4.1. General prognostic model

The algebraic model presented in Section 3 is based on the steady state versions of the energy- and flux-budget equations, Eqs. (24), (30), (34), (37); and, as any other algebraic closure, has a limited area of application (in particular, it erroneously prescribes total decay of turbulence in the regions of flow with zero mean shear, e.g., at the axes of jets). In its general form, the EFB closure employs prognostic versions of the above equations, with the non-local third-order transport terms $\Phi_K$, $\Phi_P$, $\Phi_i^{(\tau)}$ and $\Phi_z^{(F)}$ expressed through the conventional turbulent diffusion approximation:

$$\frac{DE_K}{Dt} - \frac{\partial}{\partial z} K_E \frac{\partial E_K}{\partial z} = -\tau_{i3} \frac{\partial U_i}{\partial z} + \beta F_z - \frac{E_K}{t_T}, \tag{87}$$

$$\frac{DE_P}{Dt} - \frac{\partial}{\partial z} K_E \frac{\partial E_P}{\partial z} = -\beta F_z - \frac{E_P}{C_P t_T}, \tag{88}$$

$$\frac{D\tau_{i3}}{Dt} - \frac{\partial}{\partial z} K_{FM} \frac{\partial \tau_{i3}}{\partial z} = -2E_z \frac{\partial U_i}{\partial z} - \frac{\tau_{i3}}{C_\tau t_T} \ (i = 1.2), \tag{89}$$

$$\frac{DF_z}{Dt} - \frac{\partial}{\partial z} K_{FH} \frac{\partial F_z}{\partial z} = -2(E_z - C_\theta E_P) \frac{\partial \Theta}{\partial z} - \frac{F_z}{C_F t_T}. \tag{90}$$

The turbulent transport coefficients: $K_E$ for the turbulent energies, and $K_{FM}$, $K_{FH}$ for the turbulent fluxes are taken proportional to the eddy viscosity $K_M$, Eq. (43):

$$K_E / C_E = K_{FM} / C_{FM} = K_{FH} / C_{FH} = E_z t_T, \qquad (91)$$

where $C_E$, $C_{FM}$ and $C_{FH}$ are dimensionless constants to be determined empirically.

Generally speaking, the vertical component of TKE $E_z$ is governed by the prognostic Equation (21) with the pressure terms $Q_{ii}$ determined through the inter-component energy exchange concept, Eqs. (48)-(49). For practical purposes we recommend a simpler approach based on the quite natural assumption that the TKE components are transported altogether. Then, given $E_K$, the vertical TKE [that appears in Eqs. (89), (90)] is determined as $E_z = A_z E_K$, where $A_z = A_z(\Pi)$ is determined by Eq. (78) with $\Pi = E_P / E_K$ based on the prognostic parameters $E_K$ and $E_P$:

$$E_z = A_z E_K, \quad A_z = A_z(\Pi) \, [Eq.(78)], \quad \Pi = \frac{E_P \, [Eq.(88)]}{E_K \, [Eq.(87)]}. \qquad (92)$$

We recall that the TKE $E_K$ and its dissipation rate $\varepsilon_K$ vary in space and time and are transported by both the mean flow and the turbulence. Hence, the turbulent dissipation time scale $t_T = E_K / \varepsilon_K$ is also transported in space and varies in time. In the steady state, its local-equilibrium value $t_{TE}$ is expressed through $E_K$, $A_z$ and $\Pi$ by Eqs. (76)-(78). Generally the equilibrium is, on the one hand, distorted due to non-steady and non-local processes and, on the other hand, re-established by the local adjustment mechanisms. Such counteractions are modelled by the relaxation equation:

$$\frac{Dt_T}{Dt} - \frac{\partial}{\partial z} K_T \frac{\partial t_T}{\partial z} = -C_R \left( \frac{t_T}{t_{TE}} - 1 \right), \quad t_{TE} = t_{TE}(E_K, A_z, \Pi) \, [Eq.(76)-(78)]. \qquad (93)$$

Here, the relaxation time is taken proportional to the local-equilibrium dissipation time scale $t_{TE}$ determined through $E_K$, $A_z$ and $\Pi$ by Eq. (76)-(78); $K_T = C_T E_z t_T$ is the same kind of turbulent exchange coefficient as $K_E$, $K_{FM}$, $K_{FH}$; $C_T$ and $C_R$ are dimensionless constants to be determined empirically.

By and large, the general EFB closure model consists of
(A) five prognostic Equations: (87)-(90), (93) for TKE $E_K$, TPE $E_P$, vertical turbulent flux of momentum $\tau_{i3}$ ($i = 1,2$), vertical turbulent flux of potential temperature $F_z$, and turbulent dissipation time scale $t_T$ [that determines the TKE dissipation rate $\varepsilon_K = E_K / t_T$];
(B) three diagnostic relations: Eq. (76) for the local-equilibrium turbulent time scale $t_{TE}$, Eq. (77) for $E_K / \tau$, and Eq. (78) for the vertical share of TKE $A_z$.

In addition to empirical constants of the algebraic version of the EFB closure (already determined in Section 3), the general EFB closure includes additional constants $C_E$, $C_{FM}$, $C_{FH}$, $C_T$ and $C_R$ that are to be determined through case studies by fitting results from numerical modelling with observational and LES data.

Compared to the currently used closure models, the EFB closure benefits from the following advancements:
- consistent energetics based on the prognostic budget equations for TKE $E_K$ and TPE $E_P$, Eq. (87) and (88), and reliable stratification parameter $\Pi = E_P / E_K$;
- generally non-gradient concept of the turbulent transport based on the budget equations for the turbulent fluxes, Eqs. (89)-(90);
- advanced concept of the inter-component exchange of TKE, Eqs. (48)-(50), (78);
- advanced concept of the turbulent dissipation time scale, Eqs. (76), (93).

### 4.2. Down-gradient transport models

In a number of problems the steady-state version of the flux-budget Equations (89)-(90) provides quite sufficient approximation. It essentially simplifies the model, keeping in force all the above benefits, except for the possibility to reproduce presumably rare cases of the non-gradient turbulent transports. Therefore, for massive environmental-modelling applications, the EFB closure can be reduced to the following equations:

(A) Prognostic energy budget equations, Eqs. (87) and (88), for TKE $E_K$ and TPE $E_P$, supplemented with diagnostic formulation, Eq. (92), for the vertical TKE $E_z$;

(B) Prognostic formulation, Eqs. (76)-(78), (93), for the turbulent dissipation time scale $t_T$;

(C) Steady-state versions of the flux-budget equations, Eqs. (89) and (90), that provide diagnostic down-gradient transport formulation of the vertical turbulent fluxes in terms of the eddy viscosity $K_M$ and eddy conductivity $K_H$:

$$\tau_{i3} = -K_M \frac{\partial U_i}{\partial z}, \quad F_z = -K_H \frac{\partial \Theta}{\partial z}, \qquad (94)$$

$$K_M = 2C_\tau E_z t_T, \quad K_H = 2C_F E_z t_T \left(1 - C_\theta \frac{E_P}{E_z}\right), \qquad (95)$$

where $E_z$, $E_P$ and $t_T$ are determined through the equations listed above in paragraphs (A) and (B).

As needed, the model can be further simplified keeping only two prognostic equations, Eqs. (87) and (88), for $E_K$ and $E_P$; and determining other parameters diagnostically: $E_z$ – through Eq. (92), $t_T = t_{TE}$ – through Eqs. (76)-(78), and vertical turbulent fluxes $\tau_{i3}$ and $F_z$ – through Eqs. (94) and (95).

### 4.3. Minimal prognostic model

Until recently the common practice limited to the sole use of the TKE budget equation. Mauritsen at al. (2007) and Angevine et al. (2010) employed the TTE budget equation. Be it as it may, closure models based on only one prognostic energy budget equation inevitably miss some essential features of non-steady regimes of turbulence. Principal inaccuracy of the one-equation approach is rooted in the difference between the TPE and TKE dissipation times: $C_P t_T$ and $t_T$, respectively. Because $C_P = 0.417$ (see Section 3.3), TPE dissipates faster than TKE, which is why one particular equation (does not matter for TKE, TPE or TTE) is not sufficient to accurately reproducing turbulence energetics. With this warning, we propose the simplest prognostic version of the EFB closure model based on the TTE budget equation:

$$\frac{DE}{Dt} - \frac{\partial}{\partial z} K_E \frac{\partial E}{\partial z} = -\tau_{i3} \frac{\partial U_i}{\partial z} - \frac{E}{t_T [1 - (1 - C_P) Ri_f]}. \tag{96}$$

It is derived by adding Equations (87) and (88) and expressing the sum $E_K + E_P / C_P$ approximately through diagnostic Equations (53) and (54). Equation (96) is preferable compared to the TKE budget equation because $E$ is a conserved property (it becomes an invariant in the absence of production and dissipation) in contrast to $E_K$ that continuously feeds the potential energy $E_P$. Except for $E$, all other parameters are determined in this version of the closure diagnostically:

(A) $E_K$, $E_P$ – through Eqs. (53), (54):

$$E_K = E \frac{1 - Ri_f}{1 - (1 - C_P) Ri_f}, \quad E_P = E \frac{C_P Ri_f}{1 - (1 - C_P) Ri_f}, \tag{97}$$

(B) $A_z$ and $E_z$ – through Eq. (50c):

$$A_z = \frac{C_r (1 - 2C_0 Ri_f / R_\infty)(1 - Ri_f) - 3 Ri_f}{(1 - Ri_f)\{3 + C_r [3 - 2(1 + C_0) Ri_f / R_\infty]\}}, \quad E_z = A_z E_K, \tag{98}$$

(C) $t_T$ – through Eq. (76) rewritten in terms of $Ri_f$:

$$t_{TE} = \frac{kz}{E_K^{1/2} + C_\Omega \Omega z} \left(\frac{E_K}{\tau}\right)^{3/2} \frac{1 - Ri_f / R_\infty}{1 - R_f}, \tag{99}$$

(D) $\tau_{i3}$ and $F_z$ – through Eqs. (43), (44), (97):

$$\tau_{i3} = -K_M \frac{\partial U_i}{\partial z}, \quad K_M = 2 C_\tau E_z t_T, \tag{100}$$

$$F_z = -K_H \frac{\partial \Theta}{\partial z}, \quad K_H = 2C_F E_z t_T \left(1 - \frac{C_\theta C_P Ri_f}{(1-Ri_f) A_z}\right), \tag{101}$$

(E) $Ri_f$ – through its definition, Eq. (39):

$$Ri_f = \frac{-\beta F_z}{\tau_{i3} \partial U_i / \partial z}. \tag{102}$$

Setting the l.h.s. of Eq. (96) equal to zero, this model reduces to the steady state EFB model considered in detail in Section 3.

## 5. Conclusions

Over decades, operationally used closure models conceptually followed Kolmogorov (1941, 1942): they limited representation of turbulence energetics to the TKE budget equation and employed hypothetical expressions of the eddy viscosity and eddy conductivity of the type $K_M \sim K_H \sim E_K t_T \sim E_K^{1/2} l$. This "one energy-equation approach", originally proposed for the neutrally stratified flows (and justified for the neutral stratification), became misleading when applied to the stably stratified flows. It disregarded the energy exchange between TKE and TPE controlled by the buoyancy flux $\beta F_z$ and, therefore, disguised the condition that $-\beta F_z$ in the steady state cannot exceed the shear production of TKE. This confusion gave rise to erroneous but widely believed statement that the steady-state turbulence can be maintained by the velocity shear only at small gradient Richardson numbers: $Ri < Ri_c < 1$, whereas at $Ri > Ri_c$ turbulence inevitably degenerates and the flow becomes laminar.

Obukhov (1946) was the first who applied the Kolmogorov closure to the thermally stratified atmospheric surface layer. He accounted for the term $\beta F_z$ in the TKE equation [which led him to discovering the stratification length scale $L$, Eq. (41), now called the "Obukhov scale"] but in all other respects he kept the original Kolmogorov closure absolutely unchanged. In particular, he disregarded the role of the turbulent potential energy (TPE) and the TKE↔TPE energy exchange. Moreover, Obukhov preserved even the concept of the turbulence length scale $l$ as merely proportional to the height $z$, precisely as it was stated in the above cited papers of his teacher Kolmogorov. And that is in spite of the Obukhov's own discovery of the length scale $L$, which gave him all grounds to conclude that $l$ should tend to $L$ in the strongly stable stratification. It is beyond questions that his model, generalising the logarithmic wall law for the stratified flows, has made a great stride forward in the physics of turbulence, not to mention that eventually it gave rise to the famous surface-layer similarity theory (Monin and Obukhov, 1954). However, in the context of turbulence closure problem, Obukhov's model happened to be to some extent misleading. It is due to the great authority of Kolmogorov, Obukhov and their school of turbulence, that further efforts towards development of turbulence closure models for meteorological and oceanographic applications were over half a century limited to the "mechanical closures" based on the sole use of the TKE budget equation, disregarding the TPE, and dared for only cautious corrections to Eq. (2): $K_M \sim K_H \sim E_K^{1/2} l$. This historical remark explains why rather simple "mechanical and thermodynamic" EFB turbulence closure was not developed already long ago.

Our work on the EFB closure, started from Elperin et al. (2005) and reflected in Zilitinkevich et al. (2007, 2008, 2009, 2010), has been inspired by numerous experimental and numerical modelling studies disclosed essential features of the stably stratified turbulence dramatically contradicting traditional closure models (e.g. Figure 5 demonstrating asymptotically linear *Ri*-dependence of the turbulent Prandtl number). The present paper summarises results from this work. Compared to prior versions of the EFB closure, we now advanced the concept of the inter-component exchange of TKE (Section 3.2); clarified physical meaning of the turbulent dissipation time and length scales and developed diagnostic and prognostic models for these scales (Sections 3.4 and 4.1); and formulated a hierarchy of the EFB turbulence closures of different levels of complexity designed for different applications.

The steady-state version of the EFB closure allows us to determine the stability dependencies of the velocity and temperature gradients, the eddy viscosity and eddy conductivity, and many other parameters of turbulence as functions of the dimensionless height *z/L* (Section 3.5). It sheds new light on the Monin-Obukhov (1954) and Nieuwstadt (1984) similarity theories and extends them to a much wider range of the stably stratified flows. Equation (82), linking *z/L* with the gradient Richardson number *Ri*, reveals that the notion "strongly stable stratification" is currently used in a rather uncertain sense. In boundary-layer meteorology, it implies nothing but the strongest stratifications achievable in the atmospheric boundary layer, which factually corresponds to the values of *z/L* in the interval 1 < *z/L* < 10. However, as follows from Eq. (82), *z/L* <10 corresponds to *Ri* < 1, that is to only *weakly-stable* stratification inherent to the strong-turbulence regime. On the contrary, the *strongly-stable* stratification inherent to the weak-turbulence regime is observed only outside boundary layers, in the free atmosphere, where *Ri* varies typically from 1 to $10^2$, and could peak at $10^3$ in the capping inversions above the long-lived stable boundary layers. The above terminological confusion has led to erroneous treatment of the so-called *z*-less stratification regime (associated with maximal *z/L* achievable in the surface layer) as the ultimate strongly-stable stratification regime. As a result, the similarity theory in its traditional form happened to be incapable to correctly determine the asymptotic behaviour of the similarity functions at very large *z/L*. Equations (70) and (86) refine traditional surface-layer flux-profile relationships and offer scope for improving the surface-flux algorithms in atmospheric models.

Empirical validation of a turbulence closure model often reduces to comparison with empirical data of the model results related only to the turbulent fluxes ($\boldsymbol{\tau}$, $F_z$, etc.) and the mean flow parameters ($\mathbf{U}$, $\Theta$, etc.), with no care of other conclusions from the model. Thus, we never met in literature verifications of the operationally used TKE-budget closure models in terms of the stability dependences of the ratios $E_K/\tau$ or $\varepsilon_K/(\tau S)$ in the steady state. Contrastingly, we verify results from our theory related to all the considered characteristics of turbulence, first of all, in the steady-state regime of turbulence. This work faces essential difficulties because of lack of data on the steady-state turbulence in strongly stable stratification. We were forced to very carefully select appropriate data presented in our figures. Comprehensive empirical validation of the EFB turbulence closure is yet to be performed. New, specially designed DNS and laboratory experiments are needed to realistically reproduce the weak-turbulence regime in the stationary and homogeneous conditions. Alternative validation tools provide case studies of the very stably stratified turbulent flows in the atmosphere and hydrosphere using numerical models equipped with the EFB turbulence closure employing our tentative estimates of the empirical constants.

We propose different prognostic versions of the EFB closure, from the most general (Section 4.1) to the minimal (Section 4.3), for use in different applications depending on available computational resources and scientific or operational goals. The general and the down-gradient transport versions of the EFB closure (Sections 4.1 and 4.2) are recommended for modelling the so-called "optical turbulence". The latter is controlled by the temperature-fluctuation "energy" $E_\theta = (N/\beta)^2 E_P$ (Lascaux et al., 2009) and, therefore, can not be reliably recovered from the turbulence closures disregarding the TPE budget equation. For operational numerical weather prediction, air quality and climate modelling, we recommend, as sufficiently accurate and not too computationally expensive, the three-equation version of the TKE closure (Section 4.2).

**Acknowledgements:** This work has been supported by the EC FP7 ERC Grant No. 227915 "Atmospheric planetary boundary layers – physics, modelling and role in Earth system"; the Russian Federation Government Grant No. 11.G34.31.0048 "Air-sea/land interaction: physics and observation of planetary boundary layers and quality of environment"; the Israel Science Foundation governed by the Israeli Academy of Sciences, Grants No. 259/07 and No. 1037/11; and the Norwegian Research Council Grant No. 191516/V30 "Planetary Boundary Layer Feedback in the Earth's Climate System". Our thanks to Rostislav Kouznetsov (A.M. Obukhov Institute of Atmospheric Physics, Moscow / Finnish Meteorological Institute, Helsinki) for his contribution to Figures 1-3 and 6; and to Frank Beyrich (German Weather Service) for providing us with the Lindenberg data shown in Figure 6.

# References

Angevine, W.M., Jiang H., and Mauritsen, T., 2010: Performance of an eddy diffusivity–mass flux scheme for shallow cumulus boundary layers. *Monthly Weather Rev.*, **138**, 2895–2912.

Banta, R. M., Newsom, R.K., Lundquist, J. K., Pichugina, Y. L., Coulter, R. L., and Mahrt, L., 2002: Nocturnal low-level jet characteristics over Kansas during CASES-99. *Boundary-Layer Meteorol.* **105**, 221–252.

Bertin, F., Barat, J., and Wilson, R., 1997: Energy dissipation rates, eddy diffusivity, and the Prandtl number: An in situ experimental approach and its consequences on radar estimate of turbulent parameters. *Radio Science*, **32**, 791-804.

Blackadar, A.K., 1962: The vertical distribution of wind and turbulent exchange in a neutral atmosphere. *J. Geophys. Res.*, **67**, 3095-3102.

Canuto, V.M., 2002: Critical Richardson numbers and gravity waves. *Astronomy & Astrophysics* **384**, 1119-1123

Canuto, V.M., 2009: Turbulence in astrophysical and geophysical flows. *Lect. Notes Phys.* **756**, 107–160.

Canuto, V. M., Howard, A., Cheng, Y. and Dubovikov, M. S., 2001: Ocean turbulence. Part I: One-point closure model - momentum and heat vertical diffusivities. *J. Phys. Oceanogr.*, **31**, 1413-1426.

Canuto, V.M., Cheng, Y. and Howard, A.M., 2005: What causes the divergences in local second-order closure models? *J. Atmos. Sci.* **62**, 1645-1651.

Canuto, V.M., Cheng, Y. and Howard, A.M., Esau, I.N., 2008: Stably stratified flows: A model with no Ri(cr). *J. Atmos. Sci.* **65**, 2437-2447.

Cheng, Y., Canuto, V. M. and Howard, A. M., 2002: An improved model for the turbulent PBL. *J. Atmosph. Sci.*, **59**, 1550-1565.

Churchill, S.W., 2002: A reinterpretation of the turbulent Prandtl number. *Ind. Eng. Chem. Res.* **41**, 6393 -6401.

Elperin, T., Kleeorin, N. and Rogachevskii, I., 1996: Isotropic and anisotropic spectra of passive scalar fluctuations in turbulent fluid flow. *Phys. Rev. E* **53**, 3431-3441.

Elperin, T., Kleeorin, N., Rogachevskii, I., and Zilitinkevich, S., 2002: Formation of large-scale semi-organized structures in turbulent convection. *Phys. Rev. E* **66**, 066305 (1-15).

Elperin, T., Kleeorin, N., Rogachevskii I., and Zilitinkevich, S., 2005: New turbulence closure equations for stable boundary layer. Return to Kolmogorov (1941). *5$^{th}$ Annual Meeting of the European Meteorological Society* (Utrecht, Netherlands, September 12-16, 2005), paper No. 0553.

Elperin, T., Kleeorin, N., Rogachevskii, I., and Zilitinkevich, S., 2006: Turbulence and coherent structures in geophysical convection. *Boundary-Layer Meteorol.* **119**, 449-472.

Engelbart, D.A.M., Andersson, S., Görsdorf, U., and Petenko, I. V., 2000: The Lindenberg SODAR/RASS experiment LINEX-2000: concept and first results. *Proc. 10$^{th}$ Int. Symp. Acoust. Rem. Sens.*, Auckland, New Zealand, 270-273.

Esau, I., 2004: Simulation of Ekman boundary layers by large eddy model with dynamic mixed sub-filter closure. *Environmental Fluid Mech.*, **4**, 273-303.

Esau, I., 2009: Large-eddy simulations of geophysical turbulent flows with applications to planetary boundary layer research, arXiv:0907.0103v1 (DATABASE64 could be found on ftp://ftp.nersc.no/igor/NEW%20DATABASE64/)

Esau, I. N., and Zilitinkevich, S. S., 2006: Universal dependences between turbulent and mean flow parameters in stably and neutrally stratified planetary boundary layers. *Nonlin. Processes Geophys*. **13**, 135–144.

Foken, T., 2006: 50 years of the Monin–Obukhov similarity theory. *Boundary-Layer Meteorol.* **119**, 431-447.

Garratt J.R., 1992: *The Atmospheric Boundary Layer*. Cambridge University Press, 316 pp.

Glazunov A.V., 2010: On the effect that the direction of geostrophic wind has on turbulence and quasi-ordered large-eddy structures in the atmospheric boundary layer. *Izvestiya RAN, FAO* **46**, 786-807.

Holton, J. R., 2004: *An Introduction to Dynamic Meteorology*. Academic Press, New York, 535 pp.

Itsweire, E.C., Helland, K.N., and Van Atta, C.W., 1986: The evolution of grid-generated turbulence in a stably stratified fluid. *J. Fluid Mech*. **162**, 299–338.

Kaimal, J. C., and Fennigan, J. J., 1994: *Atmospheric Boundary Layer Flows*. Oxford University Press, New York, 289 pp.

Kraus E. B., and Businger, J. A., 1994: *Atmosphere-Ocean Interaction.* Oxford University Press, Oxford and New York. 362 pp.

Kolmogorov, A. N., 1941: Energy dissipation in locally isotropic turbulence. *Doklady AN SSSR* **32**, No.1, 19-21.

Kolmogorov A.N., 1942: Equations of turbulent motion in an incompressible fluid. *Izv. AN SSSR, Ser. Fiz*. **6**, No. 1-2, 56-58.

Kondo, J., Kanechika, O., and Yasuda, N. 1978 Heat and momentum transfer under strong stability in the atmospheric surface layer. *J. Atmos. Sci.*, **35**, 1012-1021.

Kurbatsky, A. F., 2000: *Lectures on Turbulence*. Novosibirsk State University Press, Novosibirsk.

Kurbatsky, A. F., and Kurbatskaya, L. I., 2006: Three-parameter model of turbulence for the atmospheric boundary layer over an urbanized surface. *Izvestiya RAN, FAO* **42**, 439-455.

Kurbatsky, A. F., and Kurbatskaya, L. I., 2009: $E - \varepsilon - \langle \theta^2 \rangle$ turbulence closure model for an atmospheric boundary layer including the urban canopy. *Meteorol. Atmos. Phys*. **104**, 63-81.

Kurbatsky, A. F., and Kurbatskaya, L. I., 2010: On the turbulent Prandtl number in a stably stratified atmospheric boundary layer. *Izvestiya RAN, FAO* **40**, 169-177.

Lascaux, F., Masciardi, E., Hagelin, S., and Stoesz, J., 2009: Mesoscale optical turbulence simulations at Dome C. I: Surface layer thickness and seeing in the free atmosphere. *MNRAS*, 398, 849, 193.

Lorenz, E.N., 1955: Available potential energy and the maintenance of the general circulation. *Tellus* **7**, 157-167.

Lumley J.L., and Panofsky, H.A., 1964: *The Structure of Atmospheric Turbulence*. Interscience, New York, 239 pp.

L'vov, V. S., Procaccia, I. and Rudenko, O., 2008: Turbulent fluxes in stably stratified boundary layers. *Physica Scripta* **T132**, 014010, 1-15.

L'vov, V. S., Procaccia, I., and Rudenko, O., 2009: Energy conservation and second-order statistics in stably stratified turbulent boundary layers. *Env. Fluid Mech*. **9**, 267-295.

Mahrt, L., and Vickers, D., 2005: Boundary layer adjustment over small-scale changes of surface heat flux. *Boundary-Layer Meteorol*. **116**, 313-330.

Mauritsen, T., Svensson, G., Zilitinkevich, S.S., Esau, I., Enger, L., and Grisogono, B., 2007: A total turbulent energy closure model for neutrally and stably stratified atmospheric boundary layers. *J. Atmos. Sci.* **64**, 4117-4130.

Mellor, G. L., and Yamada, T., 1974: A hierarchy of turbulence closure models for planetary boundary layers. *J. Atmos. Sci*. **31**, 1791-1806.

Monin, A. S., and Obukhov, A. M., 1954: Main characteristics of the turbulent mixing in the atmospheric surface layer, *Trudy Geophys. Inst. AN. SSSR,* 24(151), 153-187.

Monin, A. S. and Yaglom, A. M., 1971: *Statistical Fluid Mechanics*. Volume 1. MIT Press, Cambridge, Massachusetts, 769 pp.

Mydlarski, L., 2003: Mixed velocity-passive scalar statistics in high-Reynolds-number turbulence. J. Fluid Mech. **475**, 173–203.

Nieuwstadt F.T.M., 1984: The turbulent structure of the stable, nocturnal boundary layer. *J. Atmos. Sci*. **41**, 2202-2216

Obukhov A.A., 2942: On the shape of the turbulent length scale in flows with arbitrary geometry. Institute of Mechanics USSR Academy of Sciences, *Applied Mathematics and Mechanics*, **6**, 209-220.

Obukhov A.M., 1946: Turbulence in thermally inhomogeneous atmosphere. *Trudy In-ta Teoret. Geofiz. AN SSSR* **1**, 95-115.

Ohya, Y., 2001: Wind-tunnel study of atmospheric stable boundary layers over a rough surface, *Boundary-Layer Meteorol.* **98**, 57-82.

Ostrovsky L.A., and Troitskaya Yu.I., 1987: A model of turbulent transfer and dynamics of turbulence in a stratified shear flow. *Izvestiya AN SSSR FAO* **23**, 1031- 1040.

Poulos, G. S., Blumen, W., Fritts, D. C., Lundquist, J. K., Sun, J., Burns, S. P., Nappo, C., Banta, R., Newsom, R., Cuxart, J., Terradellas, E., Balsley B., and Jensen, M., 2002: CASES-99: A comprehensive investigation of the stable nocturnal boundary layer, *Bull. Amer. Meteorol. Soc*. **83**, 555-581.

Rehmann, C. R., Koseff, J. R., 2004: Mean potential energy change in stratified grid turbulence. *Dynamics of Atmospheres and Oceans* **37**, 271–294.

Richardson, L. F., 1920: The supply of energy from and to atmospheric eddies. *Proc. Roy. Soc. London* A **97**, 354-373.

Rogers, M.M., Mansour, N.N., and Reynolds, W.C., 1989: An algebraic model for the turbulent flux of a passive scalar. *J. Fluid Mech*. **203**, 77–101.

Rotta, J. C., 1951: Statistische theorie nichthomogener turbulenz. *Z. Physik* **129**, 547-572.

Schumann, U., and Gerz, T., 1995: Turbulent mixing in stably stratified shear flows. *J. Appl. Meteorol*. **34**, 33-48.

Shih, L.H., Koseff, J.R., Ferziger, J.H., and Rehmann, C.R., 2000: Scaling and parameterisation of stratified homogeneous turbulent shear flow. *J. Fluid Mech*. **412**, 1–20.

Sorbjan, Z., 1989: *Structure of the Atmospheric Boundary Layer*. Prentice-Hall, Englewood Cliffs, New Jersey, 317 pp.

Sorbjan Z., 2012: A study of the stable boundary layer based on a single-column K-theory model. *Boundary-Layer Meteorol*. **142**, 33-53.

Srivat, A., and Warhaft, Z., 1983: The effect of a passive cross-stream temperature gradient on the evolution of temperature variance and the heat flux in grid turbulence. *J. Fluid Mech*. **128**, 323–346.

Strang, E.J., and Fernando, H.J.S., 2001: Vertical mixing and transports through a stratified shear layer. *J. Phys. Oceanogr*. **31**, 2026-2048.

Stretch, D.D., Rottman, J. W., Nomura, K. K., and Venayagamoorthy, S. K., 2001: Transient mixing events in stably stratified turbulence, In: *14$^{th}$ Australasian Fluid Mechanics Conference*, Adelaide, Australia, 10-14 December 2001.

Stretch, D.D., Rottman, J. W., Nomura, K. K., and Venayagamoorthy, S. K., 2001: Transient mixing events in stably stratified turbulence, In: *14$^{th}$ Australasian Fluid Mechanics Conference*, Adelaide, Australia, 10-14 December 2001.

Sukoriansky, S., and Galperin, B., 2008: Anisotropic turbulence and internal waves in stably stratified flows (QNSE theory), *Physica Scripta* **T132**, 014036, 1–8.

Tennekes, H., and Lumley, J.L., 1972: *A First Course in Turbulence*. The MIT Press, Cambridge, Massachusetts, and London, England, 300 pp.

Uttal, T., Curry, J. A., McPhee, M. G., Perovich, D. K. and 24 other co-authors, 2002: Surface Heat Budget of the Arctic Ocean. *Bull. Amer. Meteorol. Soc*. **83**, 255-276.

Venayagamoorthy, S.K., and Stretch, D.D., 2006: Lagrangian mixing in decaying stably stratified turbulence. *J. Fluid Mech*. **564**, 197–226.

Venayagamoorthy, S.V., and Stretch, D.D., 2010: On the turbulent Prandtl number inhomogeneous stably stratified turbulence. *J. Fluid Mech*. **644**, 359–369.

Yoon, K.H., and Warhaft, Z., 1990: The evolution of grid-generated turbulence under conditions of stable thermal stratification. *J. Fluid Mech*. **215**, 601–638.

Zilitinkevich, S.S., 2010: Comments on numerical simulation of homogeneous stably stratified turbulence. *Boundary-Layer Meteorol*. **136**, 161-164.

Zilitinkevich, S.S., Elperin, T., Kleeorin, N., and Rogachevskii, I., 2007: Energy- and flux-budget (EFB) turbulence closure model for the stably stratified flows. Part I: Steady-state, homogeneous regimes. *Boundary-Layer Meteorol.* **125**, 167-192.

Zilitinkevich, S.S., Elperin, T., Kleeorin, N., Rogachevskii, I., Esau, I., Mauritsen, T., and Miles, M. W., 2008: Turbulence energetics in stably stratified geophysical flows: strong and weak mixing regimes. *Quart. J. Roy. Met. Soc*. **134**, 793-799.

Zilitinkevich, S.S., Elperin, T., Kleeorin, N., L'vov, V., and Rogachevskii, I., 2009: Energy- and flux-budget (EFB) turbulence closure model for stably stratified flows. Part II: The role of internal gravity waves. *Boundary-Layer Meteorol*. **133**, 139-164.

Zilitinkevich, S.S., Esau, I., Kleeorin, N., Rogachevskii, I., and Kouznetsov, R.D., 2010: On the velocity gradient in the stably stratified sheared flows. Part I: Asymptotic analysis and applications. *Boundary-Layer Meteorol*. **135**, 505-511.

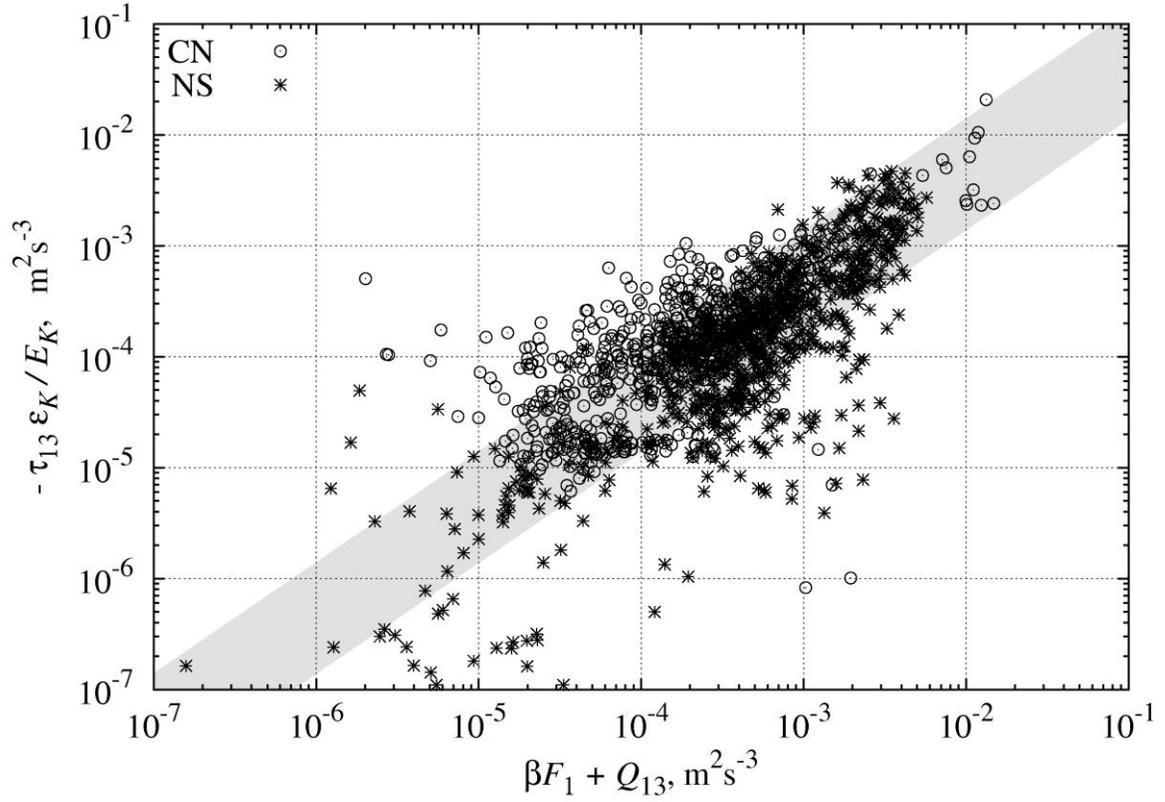

**Figure 1.** Comparison of the effective dissipation rate of the momentum flux calculated by its definition (abscissa) and by the Kolmogorov closure hypothesis (ordinate), after LES [our DATABASE64; see Esau (2004, 2009), Esau and Zilitinkevich (2007)] for conventionally neutral (CN) and nocturnal stable (NS) atmospheric boundary layers. The linear dependence (grey corridor) corresponds to our approximation, Eq. (33).

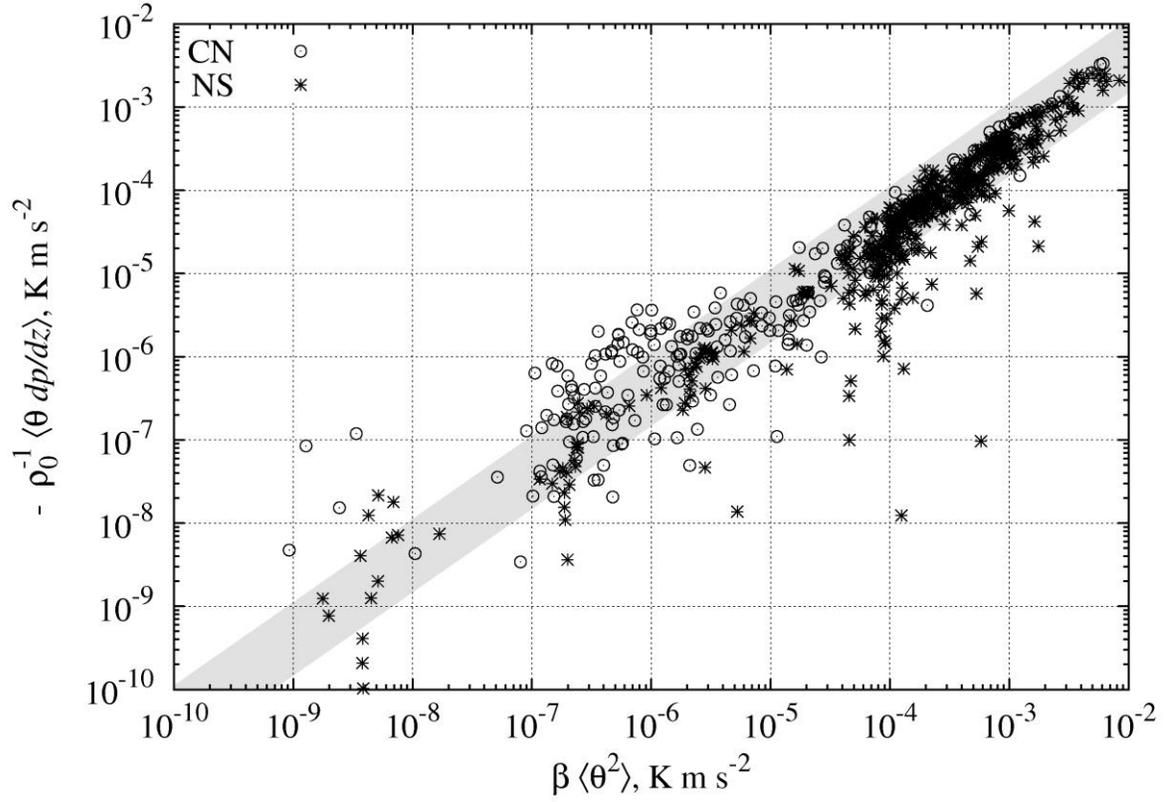

**Figure 2.** Comparison of the first (abscissa) and the second (ordinate) terms on the r.h.s. of Eq. (10), after LES (our DATABASE64). The linear dependence (grey corridor) corresponds to our approximation, Eq. (36).

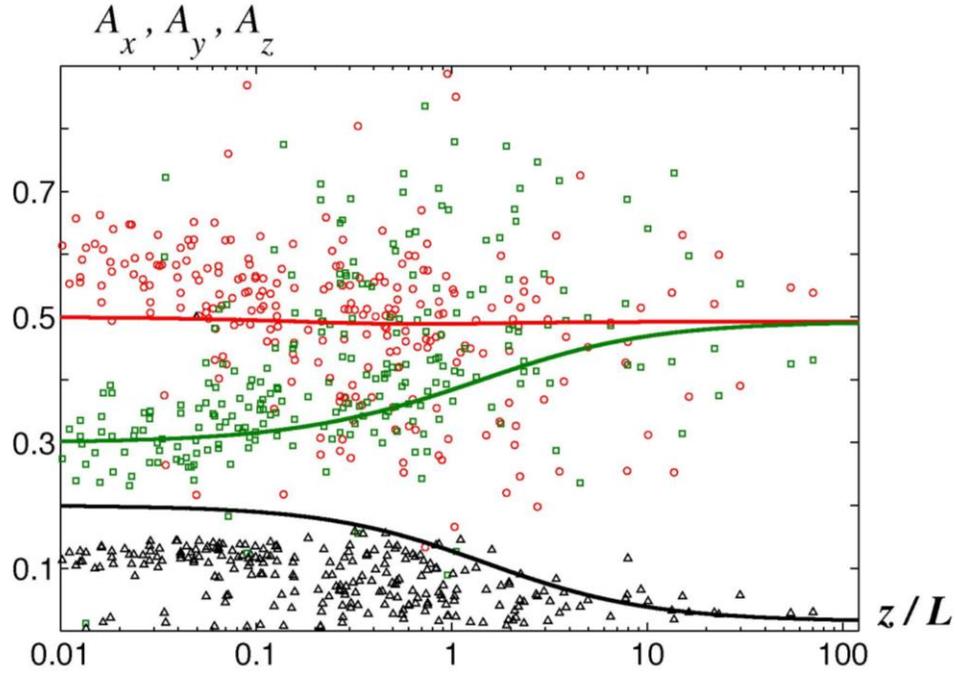

**Figure 3.** The shares of the turbulent kinetic energy $E_K$: longitudinal $A_x = E_x / E_K$ (along the mean wind, red circles), transverse $A_y = E_y / E_K$ (green squares) and vertical $A_z = E_z / E_K$ (black triangles), after Kalmykia-2007 field campaign of the A.M. Obukhov Institute of Atmospheric Physics of the Russian Academy of Sciences (courtesy of Rostislav Kouznetsov). The lines show our inter-component energy exchange model, Eq. (50), with $C_0 = 0.125$, $C_1 = 0.25$ and $C_2 = 1.01$, converted into $z/L$ dependences with the aid of Eq. (71).

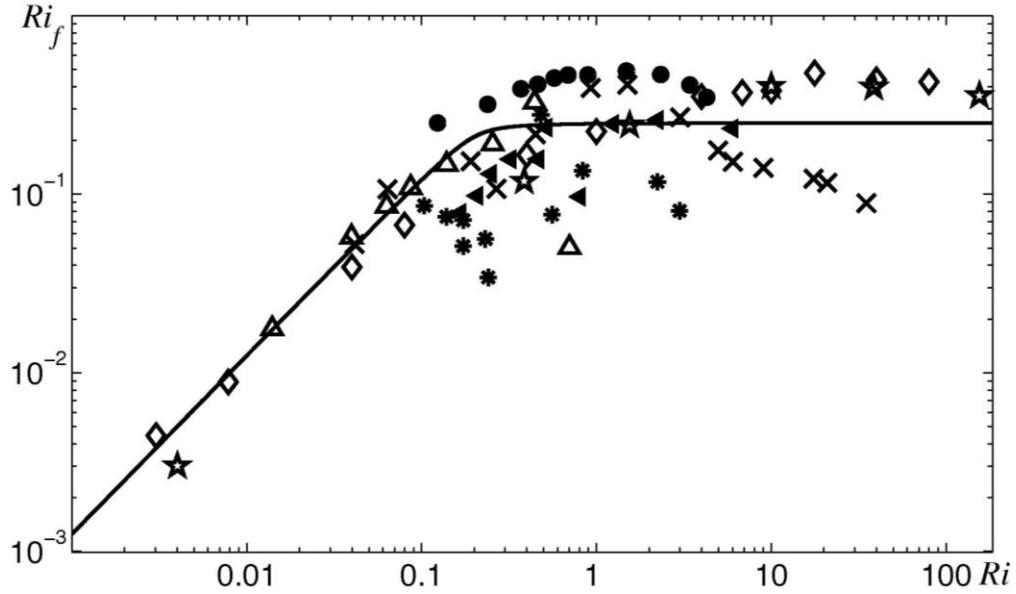

**Figure 4.** $Ri$-dependence of the flux Richardson number $Ri_f = -\beta F_z/(\tau S)$ after *meteorological observations*: slanting black triangles (Kondo et al., 1978), snowflakes (Bertin et al., 1997); *laboratory experiments*: slanting crosses (Rehmann and Koseff, 2004), diamonds (Ohya, 2001), black circles (Strang and Fernando, 2001); *DNS*: five-pointed stars (Stretch et al., 2001); *LES*: triangles (our DATABASE64). Solid line shows the steady-state EFB model, Eq. (56), with $Ri_f \to R_\infty = 0.25$ at $Ri \to \infty$.

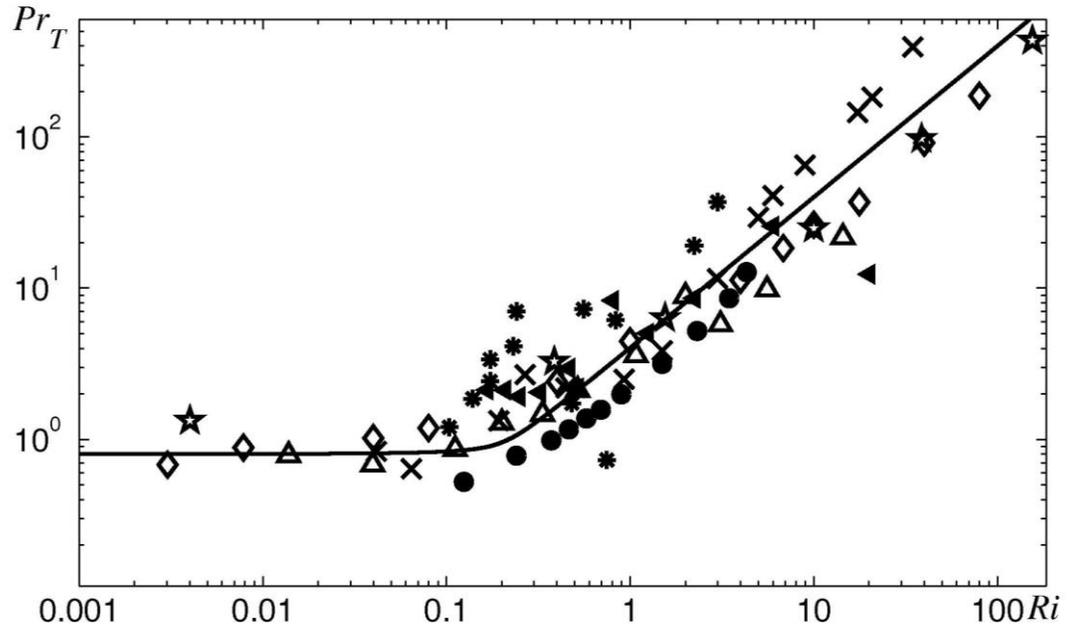

**Figure 5.** *Ri*-dependence of the turbulent Prandtl number $Pr_T = K_M / K_H$, after the same data as in Figure 4 (meteorological observations, laboratory experiments, DNS, and LES). Solid line shows the steady-state EFB model, Eq. (56).

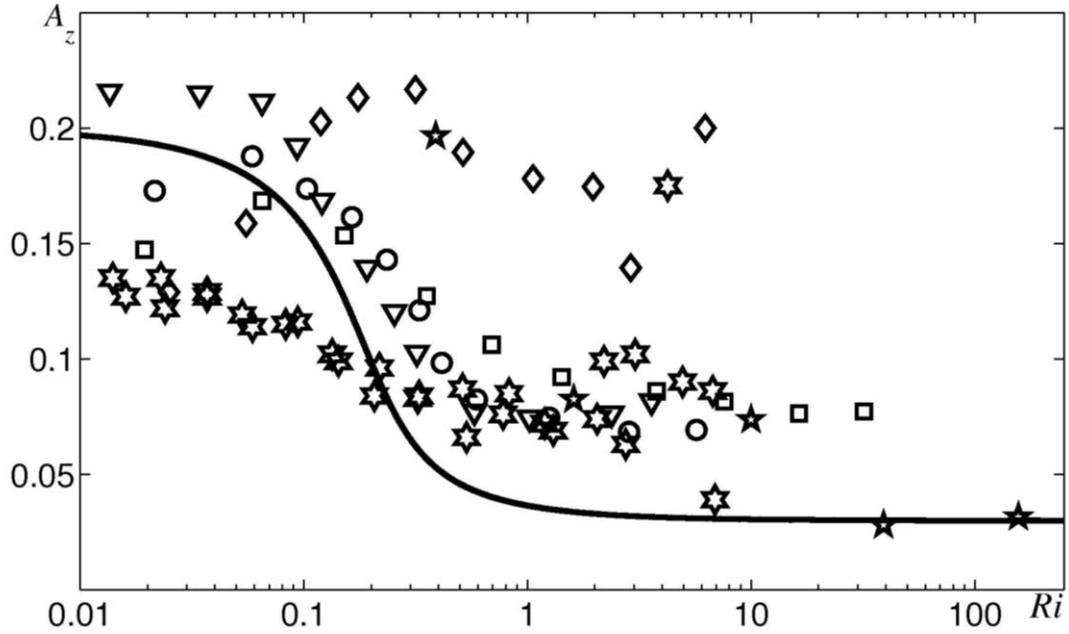

**Figure 6.** *Ri*-dependence of the vertical share of TKE $A_z = E_z / E_K$, after *meteorological observations*: squares [CME = Carbon in the Mountains Experiment, Mahrt and Vickers (2005)], circles [SHEBA = Surface Heat Budget of the Arctic Ocean, Uttal et al. (2002)], overturned triangles [CASES-99 = Cooperative Atmosphere-Surface Exchange Study, Poulos et al. (2002), Banta et al. (2002)], six-pointed stars [Lindenberg station of the German Weather Service, Engelbart et al. (2000)]; *laboratory experiments*: diamonds (Ohya, 2001); *DNS*: five-pointed stars (Stretch et al., 2001). Solid line shows the steady-state EFB model, Eqs. (50c) and (56), with $C_0 = 0.125$.

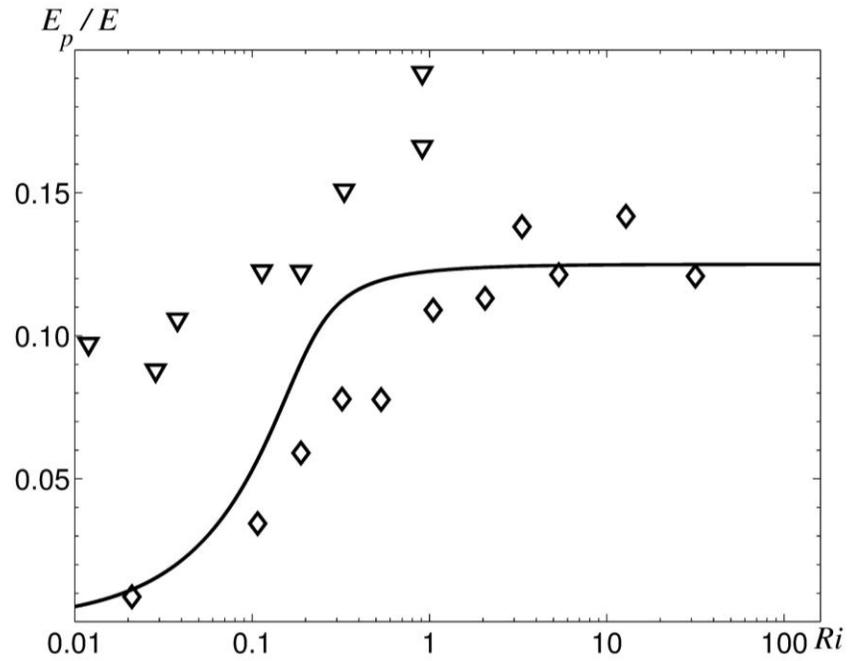

**Figure 7.** Very new figure: *Ri*-dependence of the potential-to-total turbulent energy ratio $E_P/E$, after *meteorological observations*: overturned triangles (CASES-99), *LES*: triangles (our DATABASE64), and *laboratory experiments*: diamonds (Ohya, 2001). Solid line shows the steady-state EFB model, Eqs. (54), (56).

(This Figure has been again corrected; published Figure is wrong)

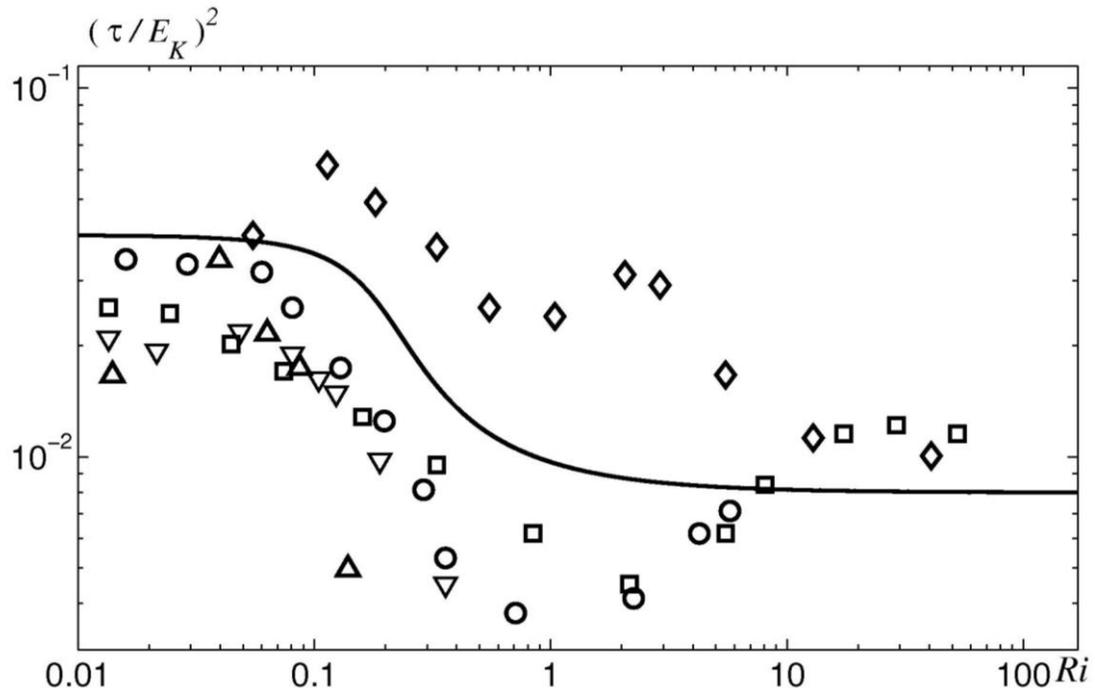

**Figure 8.** *Ri*-dependence of the squared dimensionless turbulent flux of momentum $(\tau/E_K)^2$, after *meteorological observations*: squares (CME), circles (SHEBA), overturned triangles (CASES-99); *laboratory experiments*: diamonds (Ohya, 2001); *LES*: triangles (our DATABASE64). Solid line shows the steady-state EFB model, Eq. (60).

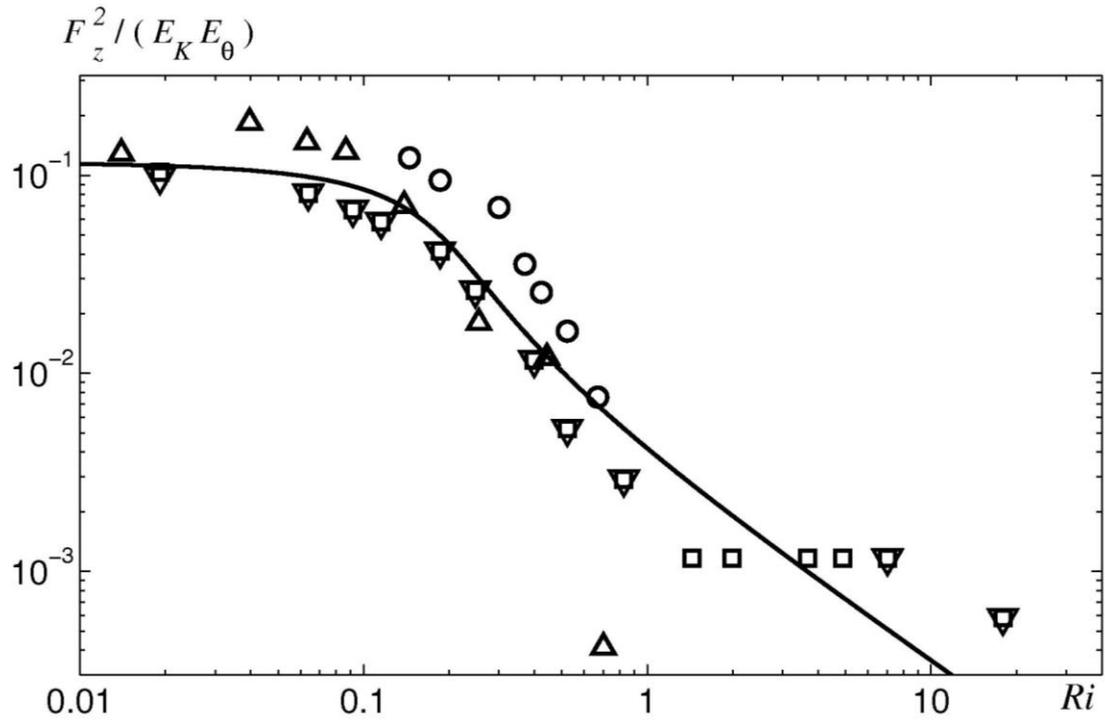

**Figure 9.** *Ri*-dependence of the squared dimensionless turbulent flux of potential temperature $F_z^2/(E_K E_\theta)$, after *meteorological observations*: squares (CME), circles (SHEBA), overturned triangles (CASES-99); *laboratory experiments*: diamonds (Ohya, 2001); LES : triangles (our DATABASE64). Solid line shows the steady-state EFB model, Eq. (61).

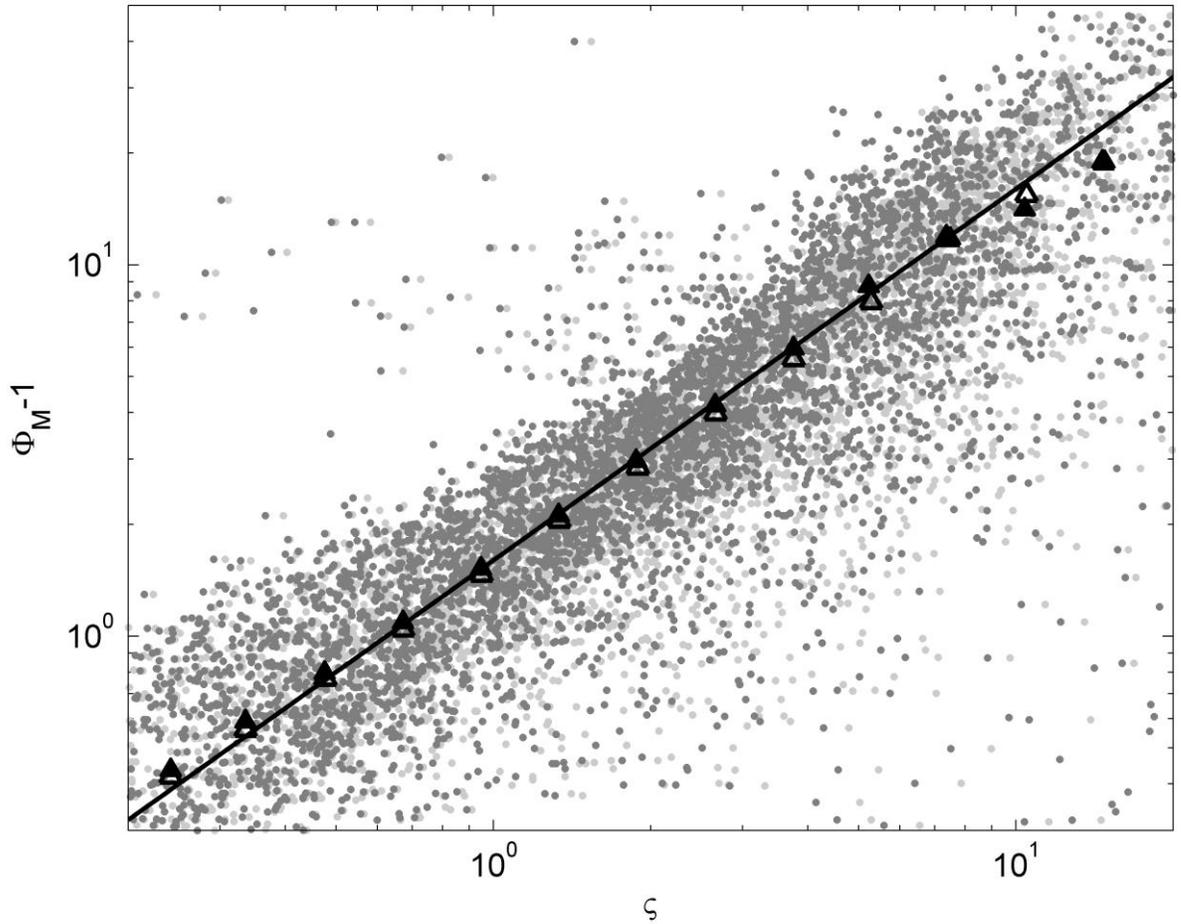

**Figure 10.** Dimensionless wind-velocity gradient $\Phi_M = (kz/\tau^{1/2})(\partial U/\partial z)$ versus dimensionless height $\varsigma$ based on the Obukhov length $L$ in the stably stratified atmospheric boundary layer, after LES (our DATABASE64). Solid line is plotted after Eq. (70) with $C_u = k/R_\infty = 1.6$. Open triangles correspond to $\varsigma = z/L$, black triangles to $\varsigma = z/[(1+C_\Omega \Omega z/E_K^{1/2})L]$ with $C_\Omega = 1$.

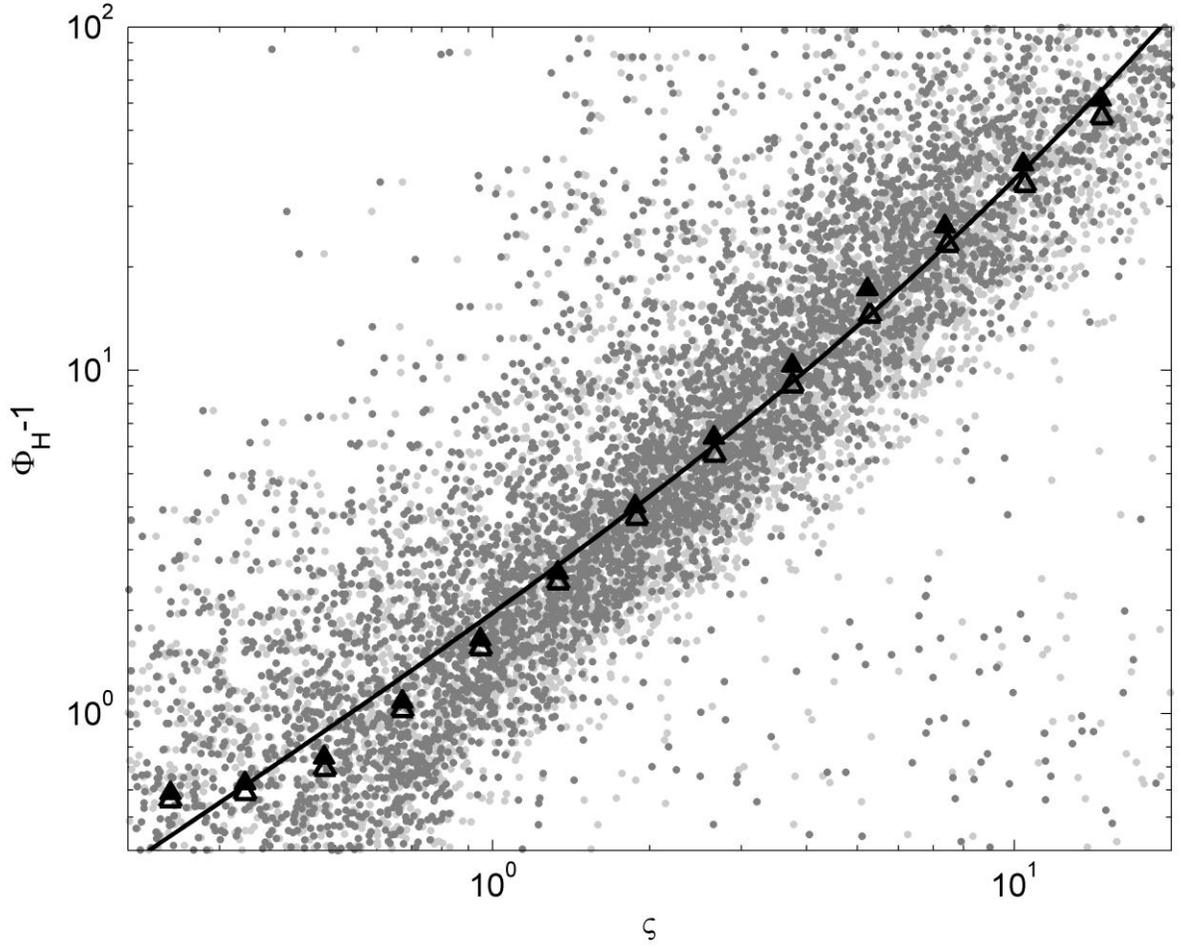

**Figure 11.** Same as in Figure 10 but for the dimensionless potential temperature gradient $\Phi_H = (-k_T z \tau^{1/2}/F_z)(\partial \Theta/\partial z)$. Solid line is plotted after Eq. (86). Open triangles correspond to $\varsigma = z/L$, black triangles to $\varsigma = z/[(1+C_\Omega \Omega z/E_K^{1/2})L]$.

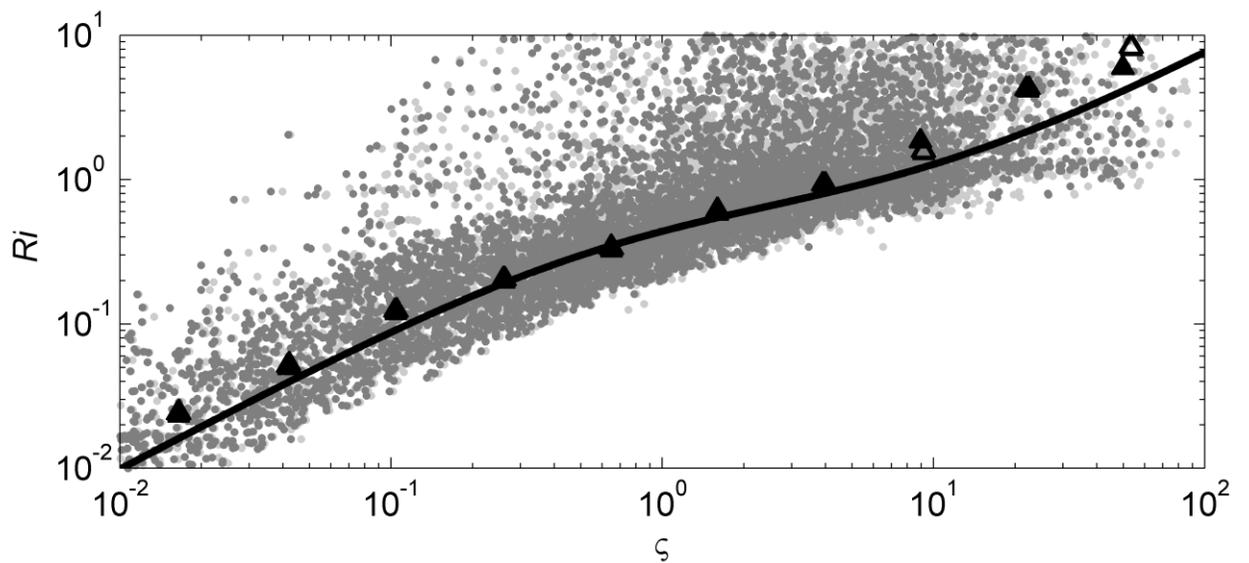

**Figure.** 12. Gradient Richardson number *Ri* versus dimensionless parameters $\varsigma = z/L$ (white triangles) and $\varsigma = l_0/L$ (black triangles) based on the Obukhov length scale *L*, after our LES. Solid line shows our model. Open triangles correspond to $\varsigma = z/L$, black triangles to $\varsigma = z/[(1+C_\Omega \Omega z / E_K^{1/2})L]$.